\documentclass[iop,numberedappendix,appendixfloats]{emulateapj}
\usepackage[dvipdfmx]{color}
\usepackage[hyperfootnotes]{hyperref}
\usepackage{color}
\newcommand{\idrop}{$i$-dropout}

\newcommand{\ydrop}{$Y$-dropout}
\newcommand{\yjdrop}{$Y\!J$-dropout}

\newcommand{\bFilter}{$B_{435}$}
\newcommand{\vFilter}{$V_{606}$}
\newcommand{\iFilter}{$i_{814}$}

\newcommand{\yFilter}{$Y_{105}$}
\newcommand{\jFilter}{$J_{125}$}
\newcommand{\jhFilter}{$J\!H_{140}$}

\newcommand{\hFilter}{$H_{160}$}

\newcommand{\clone}{Abell 2744}
\newcommand{\cltwo}{MACS J0416.1$-$2403}
\newcommand{\clthree}{MACS J0717.5+3745}
\newcommand{\clfour}{MACS J1149.6+2223}
\newcommand{\clfive}{Abell S1063}
\newcommand{\clsix}{Abell 370}

\slugcomment{Accepted for publication in ApJ}

\shorttitle{Precise mass modeling of four HFF clusters}
\shortauthors{Kawamata et al.}

\begin{document}

\title{Precise Strong Lensing Mass Modeling of Four Hubble Frontier
  Fields Clusters and a Sample of Magnified High-Redshift Galaxies}

\author{Ryota~Kawamata\altaffilmark{1},
Masamune~Oguri\altaffilmark{2,3,4}, 
Masafumi~Ishigaki\altaffilmark{2,5}, 
Kazuhiro~Shimasaku\altaffilmark{1,3},
and Masami~Ouchi\altaffilmark{4,5}
}

\altaffiltext{}{Email: kawamata@astron.s.u-tokyo.ac.jp}
\altaffiltext{1}{Department of Astronomy, Graduate School of Science, The University of Tokyo, 7-3-1 Hongo, Bunkyo-ku, Tokyo 113-0033, Japan}
\altaffiltext{2}{Department of Physics, Graduate School of Science, The University of Tokyo, 7-3-1 Hongo, Bunkyo-ku, Tokyo 113-0033, Japan}
\altaffiltext{3}{Research Center for the Early Universe, The University of Tokyo, 7-3-1 Hongo, Bunkyo-ku, Tokyo 113-0033, Japan}
\altaffiltext{4}{Kavli Institute for the Physics and Mathematics of the Universe (Kavli IPMU, WPI), The University of Tokyo, 5-1-5 Kashiwanoha, Kashiwa, Chiba 277-8583, Japan}
\altaffiltext{5}{Institute for Cosmic Ray Research, The University of Tokyo, 5-1-5 Kashiwanoha, Kashiwa, Chiba 277-8582, Japan}

\begin{abstract}
We conduct precise strong lensing mass modeling of four {\it Hubble}
Frontier Fields (HFF) clusters, Abell~2744, MACS~J0416.1$-$2403,
MACS~J0717.5+3745, and MACS~J1149.6+2223, for which HFF imaging
observations are completed.  We construct a refined sample of more than
100 multiple images for each cluster by taking advantage of the full
depth HFF images, and conduct mass modeling using 
the {\sc glafic} software, which assumes simply
parametrized mass distributions. 
Our mass modeling also exploits a magnification
constraint from the lensed Type Ia supernova HFF14Tom for
Abell~2744 and positional constraints from the multiple images S1--S4 of the
lensed supernova SN Refsdal for MACS~J1149.6+2223.
We find that our best-fitting mass models reproduce the observed image
positions with RMS errors of $\sim 0\farcs4$,
which are smaller than RMS errors in previous mass
modeling that adopted similar numbers of multiple images.
Our model predicts a new image of SN Refsdal 
with a relative time delay and magnification that are fully consistent 
with a recent detection of reappearance.
We then construct catalogs of $z\sim 6-9$ dropout galaxies behind the
four clusters and estimate magnification factors for these dropout
galaxies with our best-fitting mass models. The dropout sample from
the four cluster fields contains $\sim 120$ galaxies at $z\ga 6$,
about 20 of which are predicted to be magnified by a factor of more than
10. Some of the high-redshift galaxies detected in the HFF have
lensing-corrected magnitudes of $M_{\rm UV}\sim -15$ to $-14$. Our
analysis demonstrates that the HFF data indeed offer an ideal
opportunity to study faint high-redshift galaxies. 
All lensing maps produced from our mass modeling will 
be made available on the STScI website
\footnote{\url{https://archive.stsci.edu/prepds/frontier/lensmodels/}}.
\end{abstract}

%galaxies: photometry, galaxies: halos

\section{Introduction}
Studies of faint high-redshift galaxies can be significantly improved
by utilizing massive clusters of galaxies as natural telescopes. This
is made possible by the so-called gravitational lensing effect, in
which the propagation of a light ray is deflected by an intervening
matter distribution \citep{schneider92}. 
Although rare, extremely strong lensing events provide an opportunity 
to study very distant galaxies using their highly magnified images that
otherwise cannot even be detected.

The {\it Hubble} Frontier Fields (HFF; PI: J. Lotz) is an on-going
public {\it Hubble Space Telescope (HST)} survey to image six massive
clusters. The main purpose of the HFF is to study properties and
populations of faint high-redshift galaxies behind the cores of these
clusters with help of lensing magnifications.  Analyses of early HFF
data have already produced useful results on faint-end luminosity
functions of high-redshift galaxies
\citep{coe15,atek14,atek15,atek16,ishigaki15,oesch15,mcleod15}, size
evolution of galaxies \citep{kawamata15a}, and deep spectroscopy of
faint high-redshift galaxies \citep{vanzella14,zitrin15b}.

A key ingredient for the analysis of the HFF data is precise mass
modeling of the lensing clusters. This is because we need to convert
observed quantities, such as apparent magnitudes and angular sizes of
galaxies, to physical quantities such as intrinsic luminosities and
physical sizes which require corrections of gravitational lensing
effects. The mass distribution of the core of a cluster is usually
constrained so that it can reproduce the positions of multiple images
behind the cluster.
A lot of efforts had been made for mass
modeling before the HFF observations started, using pre-HFF data, in
order to allow prompt analyses of the HFF data by the community
\citep[e.g.,][]{richard14,johnson14,zitrin15a}.

The accuracy of mass modeling relies on 
the number of multiply imaged background galaxies. 
Much deeper {\it HST} images obtained by the HFF in
fact allow one to identify many more multiply imaged galaxies and
therefore improve strong lensing mass modeling
\citep[e.g.,][]{jauzac14,jauzac15,lam14,diego15a,diego15b,limousin15}. In addition,
spectroscopy of these multiple images is crucial for robust
identification of multiple images as well as constraining
the mass distribution, particularly the radial density profile.
Significant efforts are being made to
collect spectroscopic redshifts of galaxies detected in the HFF 
\citep[e.g.,][]{schmidt14,grillo15,karman15,wang15,treu15a,sebesta15}.

In this paper, we present our mass modeling results of the  
first four HFF clusters, \clone~\citep{abell58},
\cltwo~\citep{mann12}, \clthree~\citep{ebeling07}, and
\clfour~\citep{ebeling07}, using the full-depth HFF data as well as
the latest follow-up data. For each cluster we use more than 100
multiple images to constrain the mass distribution assuming a simply
parametrized mass model. We then construct $z\sim6-9$ dropout galaxy
catalogs in these clusters. Our mass modeling results are used to derive
magnification factors for these high-redshift galaxies. We also
discuss whether these high-redshift galaxies are multiply imaged or
not. 

The structure of our paper is as follows.
In Section~\ref{sec:data}, we describe the {\it HST} data
used in the paper, as well as the construction of photometric
catalogs. Our mass modeling procedure is described in detail in
Section~\ref{sec:model}, and the results of the mass modeling are given in
Section~\ref{sec:result}. We construct $z\sim6-9$ dropout galaxy
catalogs in Section~\ref{sec:dropout}. Finally, we summarize our
results in Section~\ref{sec:conclusion}.  Throughout this paper, 
we adopt a flat cosmological model with the matter density
$\Omega_{M} = 0.3$, the cosmological constant $\Omega_{\Lambda} = 0.7$,
and the Hubble constant $H_{0} = 70~{\rm km\,s^{-1}Mpc^{-1}}$.
Magnitudes are given in the AB system \citep{okeg83}
and coordinates are given in J2000.

\section{{\it HST} Data}\label{sec:data}
\subsection{{\it HST} Images}
We use the public HFF
data\footnote{\url{http://www.stsci.edu/hst/campaigns/frontier-fields/}}
for our analysis.
The HFF targets six massive clusters, \clone\ ($z=0.308$), \cltwo\ ($z=0.397$), 
\clthree\ ($z=0.545$), \clfour\ ($z=0.541$), 
\clfive\ ($z=0.348$), and \clsix\ ($z=0.375$), which have been chosen
according to their lensing strength and also their accessibility from major
ground-based telescopes. The cluster core and parallel field region of
each cluster are observed deeply with the IR channel of Wide Field
Camera 3 (WFC3/IR) and the Advanced Camera for Surveys (ACS).
As of October 2015, {\it HST} observations for the first four 
clusters, \clone, \cltwo, \clthree, and \clfour, are completed. 
Observations of the remaining two clusters
will be completed by mid 2016.

\begin{figure}[t]
  \centering
      \includegraphics[width=0.49\linewidth, trim=100 60 100 30, clip]{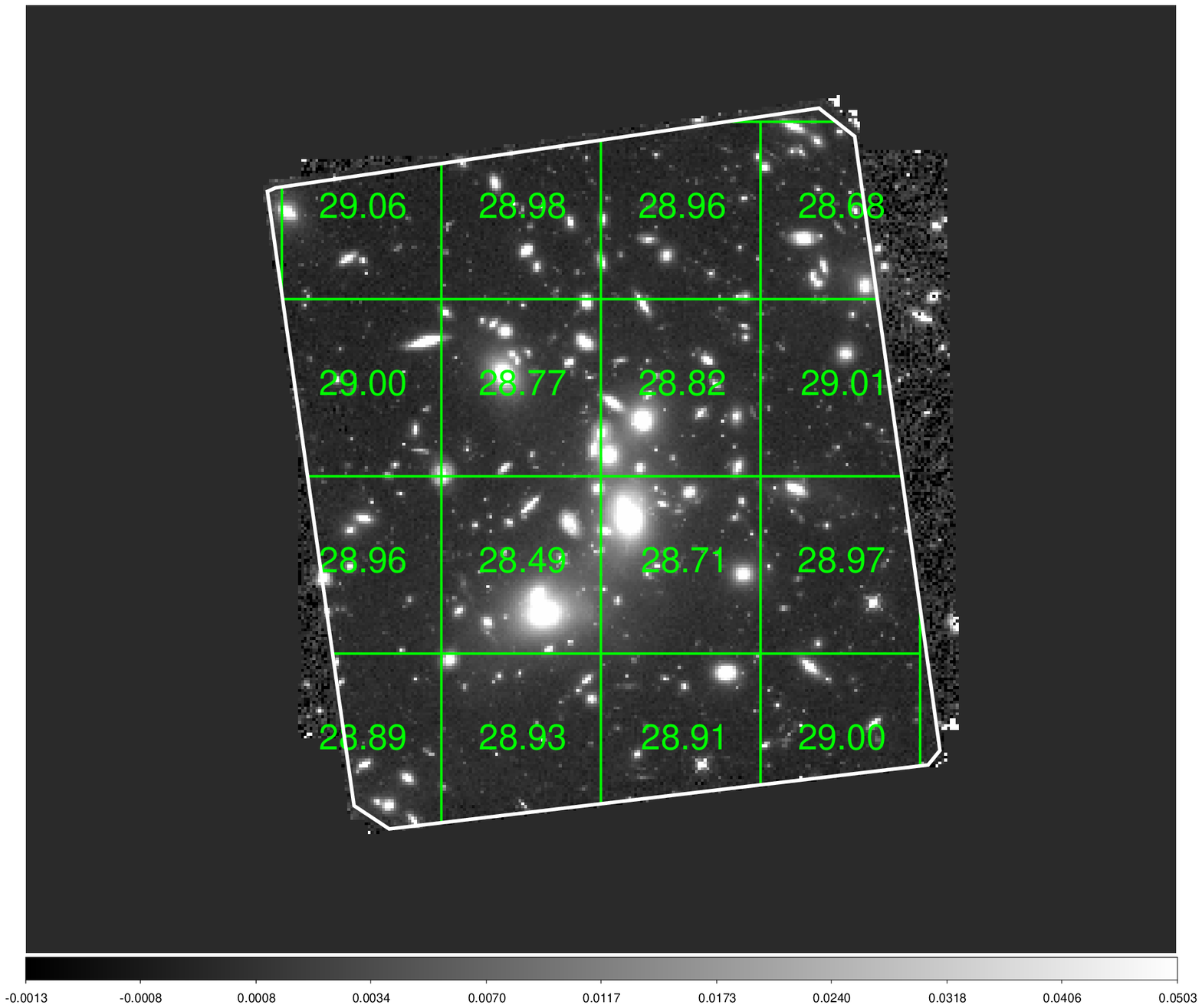}
      \includegraphics[width=0.49\linewidth, trim=100 50 100 40, clip]{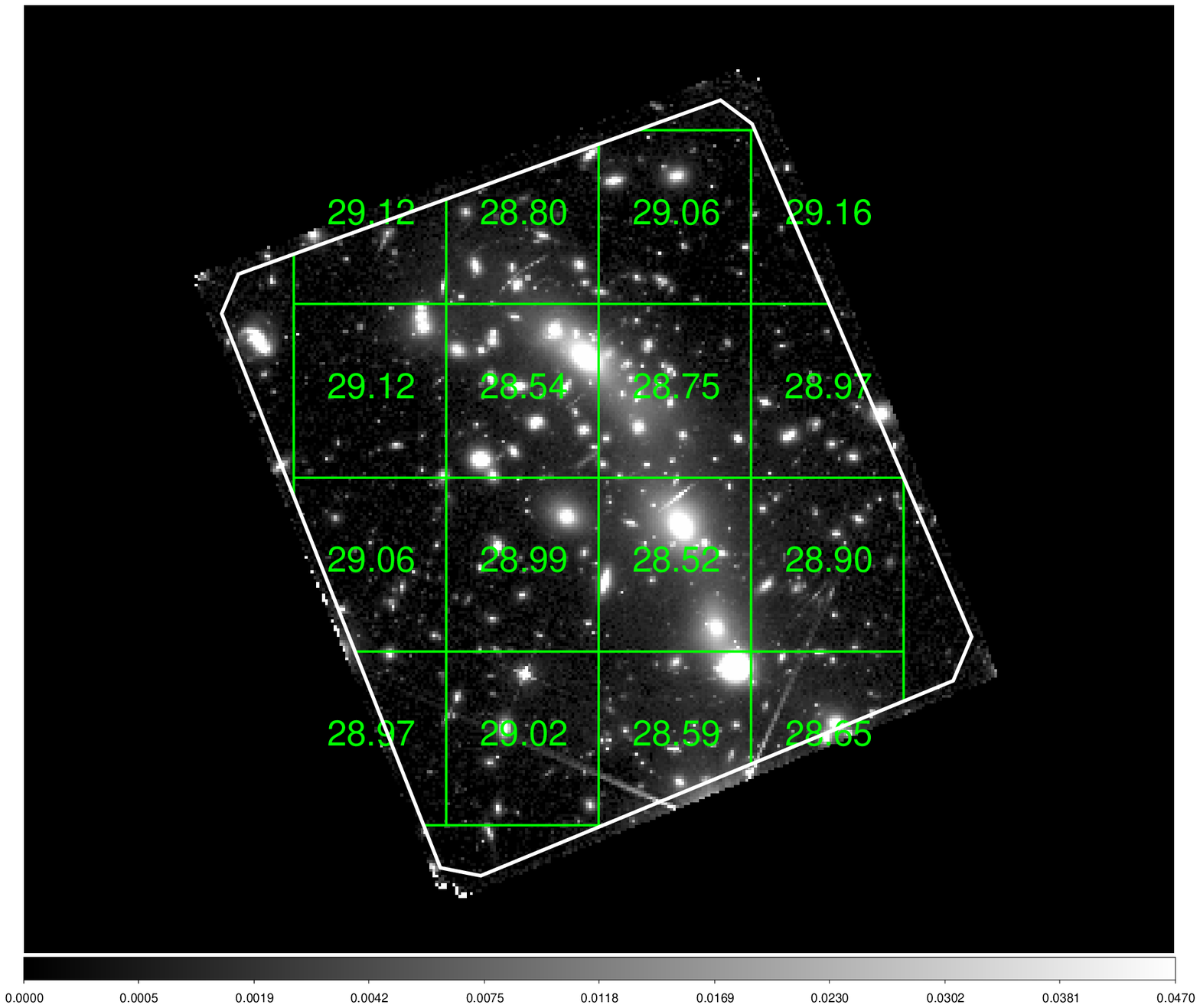}\\
      \vspace{2pt}
      \includegraphics[width=0.49\linewidth, trim=60 40 70 20, clip]{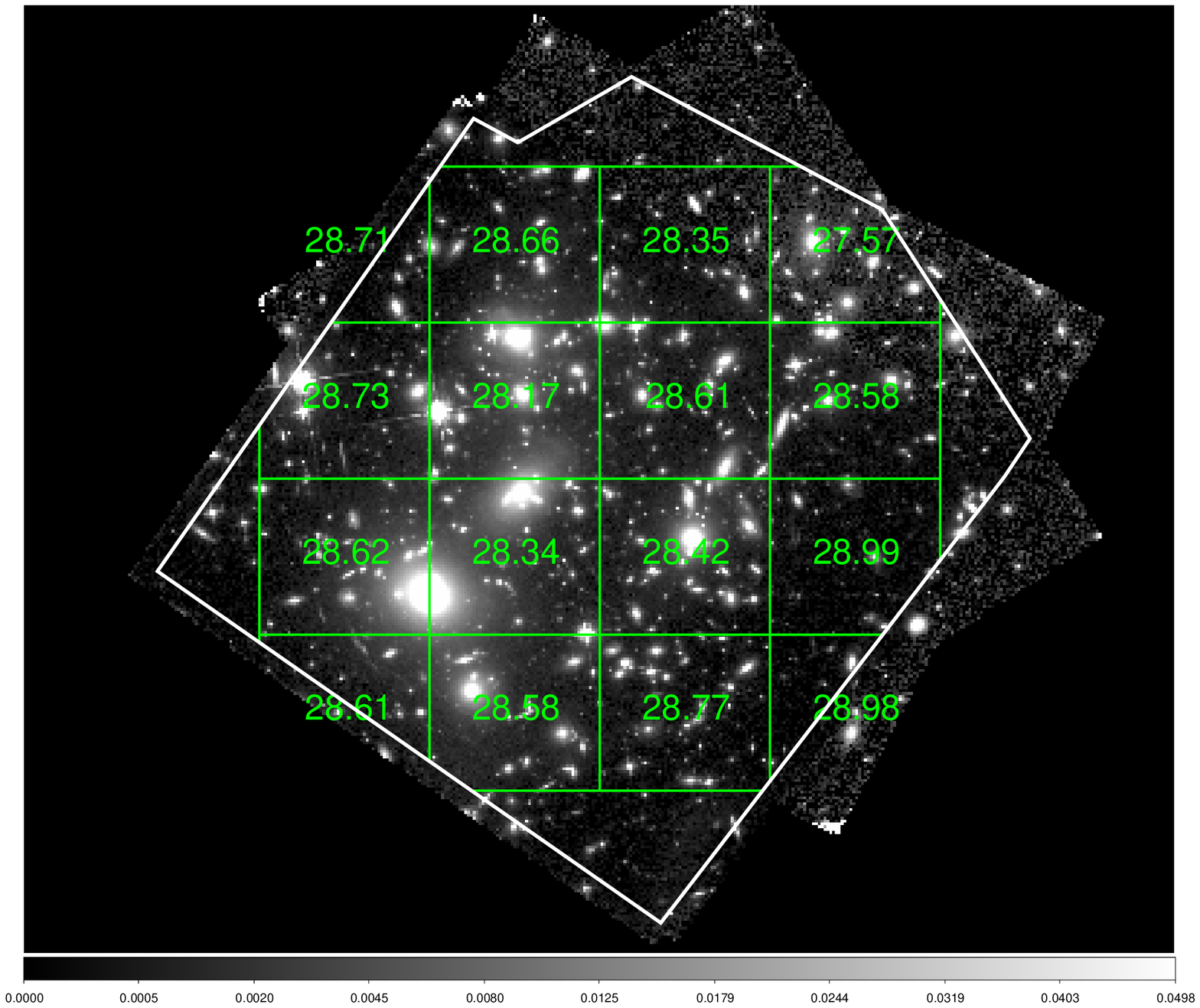}
      \includegraphics[width=0.49\linewidth, trim=90 65 90 45, clip]{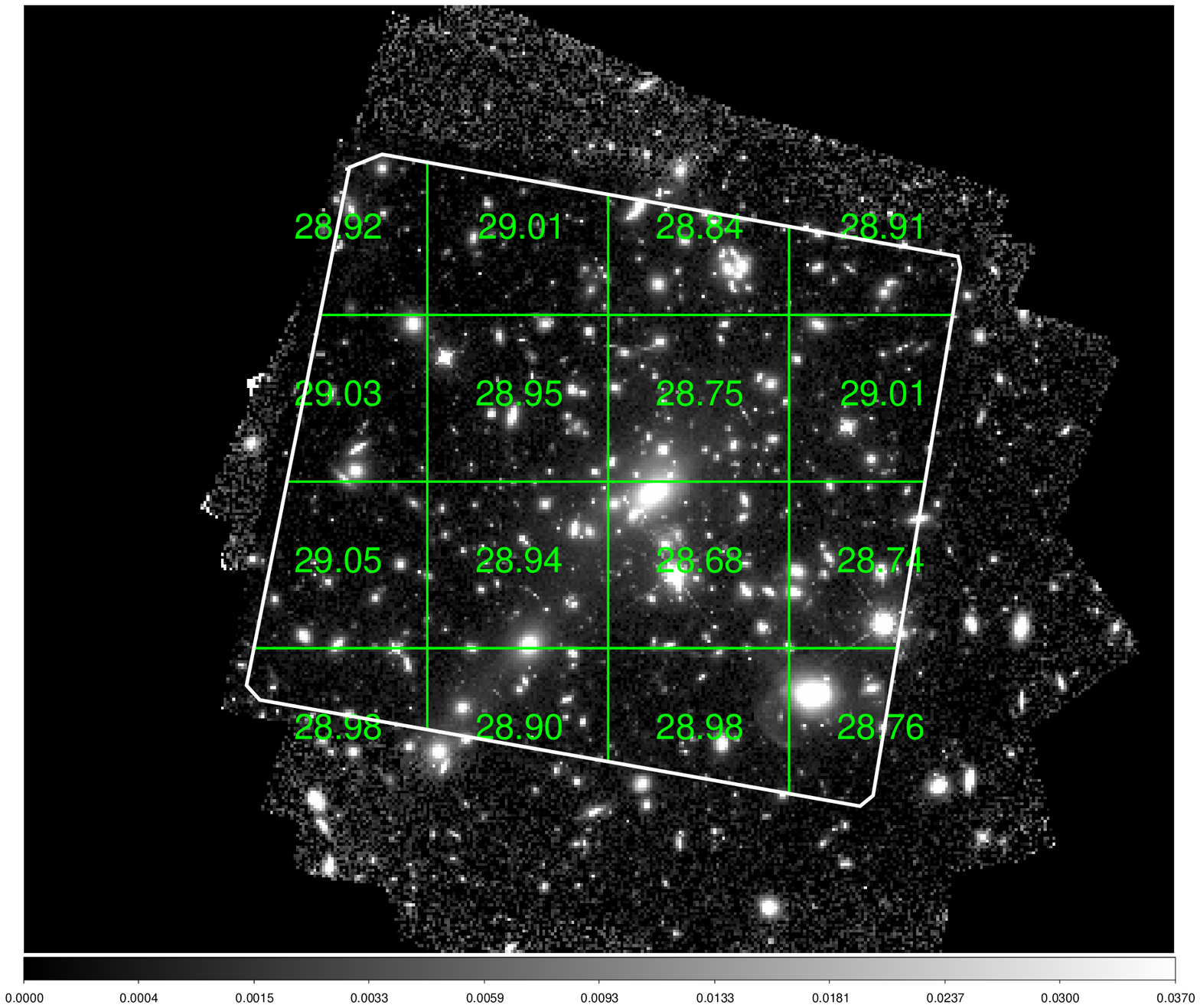}
  \caption{Measured $5\sigma$ limiting magnitudes in
      $4\times 4$ grid cells. \hFilter\ limiting magnitudes are
      shown on the {\it HST} \hFilter\ band image for 
 \clone\ ({\it upper left}), \cltwo\ ({\it upper right}),
  \clthree\ ({\it lower left}), and \clfour\ ({\it lower right}). 
  Each number shows the limiting magnitude in each cell.
  We use only the regions within the white lines to 
  search for high-redshift galaxies.
  }
  \label{fig:limitmag}
\end{figure}

In this study, we use the Version 1.0 data products of drizzled images
with a pixel scale of $0\farcs03\ {\rm pixel^{-1}} $ provided by
Space Telescope Science Institute (STScI). The images for each cluster
consist of F435W (\bFilter), F606W (\vFilter), and F814W (\iFilter)
images from ACS, and F105W (\yFilter), F125W (\jFilter), F140W
(\jhFilter), and F160W (\hFilter) images from WFC3/IR.
While we use standard correction mosaics for the ACS images,
we choose mosaics corrected for time-variable background 
sky emission for the WFC3/IR images when available.
In order to take account of the inhomogeneity of the
  limiting magnitude due to, e.g., intracluster light, we divide the
  WFC3/IR field of view of each cluster into $4\times 4$ grid cells
  and measure limiting magnitudes in individual
  cells, as shown in Figure~\ref{fig:limitmag}.
The $5\sigma$ limiting magnitudes 
on a $0\farcs35$ diameter aperture 
of these images are $\sim 29$~mag.

Three out of the four clusters have also been observed with {\it HST}
in the CLASH project \citep[see][for more details]{postman12}.
Although the CLASH imaging uses many additional bands (F225W, F275W,
F336W, F390W, F475W, F625W, F775W, F850LP, and F110W), we do not use
these images because they are considerably shallower than the HFF
images. 

\subsection{Photometric Catalogs}
We construct two different photometric catalogs specified for the
following two purposes, (1) selection of cluster member galaxies and
(2) detection of faint high-redshift galaxies,
using the method similar to the one used in
\citet{ishigaki15}. Here we briefly describe the method to construct
the photometric catalogs. 

Member galaxies are selected utilizing both the red sequence and
photometric redshift techniques. For accurate estimates of galaxy
colors, we convolve {\it HST} images with a Gaussian kernel in order
to match the point-spread function (PSF) sizes of all images of
interest to the largest one. Then, we run {\sc SExtractor}
\citep{bertin96} in dual-image mode using the \iFilter\ image as the 
detection image setting the parameters {\tt DEBLEND\_MINCONT =
  0.00005}, {\tt DEBLEND\_NTHRESH = 50}, {\tt DETECT\_MINAREA = 5},
and {\tt DETECT\_THRESH = 2.5}. 
We estimate photometric redshifts of the galaxies in this catalog
using {\sc BPZ} \citep{benitez00}.
We use the \bFilter $-$\vFilter\
color-magnitude diagram to identify the red sequence, and extract
cluster members with \vFilter -band magnitudes brighter than $\sim 24-25$~mag
\citep[see][for more details]{ishigaki15}.  We then select galaxies 
in the vicinity of the red sequence whose photometric redshifts 
coincide with the cluster redshift as cluster members.
After applying these criteria,
we refine the member galaxy catalog by adding and removing some
galaxies based on their colors, morphologies, and spectroscopic redshifts
\citep{owers11, ebeling14}.

In the construction of a photometric catalog of high-redshift
galaxies, we co-add three bands (\jFilter, \jhFilter, and \hFilter) 
for the $i$- and $Y$-dropout selection and two bands (\jhFilter\ and \hFilter) 
for the $Y\!J$-dropout selection using
{\sc SWarp} \citep{bertin02}.
Weight images of these co-added images are also produced
from public weight images.
Before running {\sc SExtractor} to build photometric
catalogs, we again match PSF sizes for reliable color measurements.
For the $i$-dropout selection, images for all the bands are
PSF-matched except for \bFilter\ and \vFilter, and for the other
selections all except for \bFilter, \vFilter, and \iFilter\ are
PSF-matched. Then, we run {\sc SExtractor} in dual-image mode using 
the co-added images as the detection image with the parameters of 
{\tt DEBLEND\_MINCONT = 0.0005}, 
{\tt DEBLEND\_NTHRESH = 16}, {\tt DETECT\_MINAREA = 4},
and {\tt DETECT\_THRESH = 3.0}.
For measuring colors of galaxies, we use aperture magnitudes ({\tt
  MAG\_APER}) $m_{\rm AP}$ with a aperture diameter of $0\farcs35$
for the convolved images and $0\farcs20$ (\bFilter), $0\farcs19$ (\vFilter), and
$0\farcs19$ (\iFilter) for the non-convolved images. 
Total magnitudes of galaxies are also derived from {\tt MAG\_APER} 
magnitudes with the aperture correction derived in \citet{ishigaki15}.
Specifically, the aperture correction factor $c_{\rm AP}$ is
$c_{\rm AP} = 0.82$, which is defined such that the total magnitude
$m_{\rm tot}$ is given by $m_{\rm tot} = m_{\rm AP} - c_{\rm AP}$.

We also derive photometric redshifts for the high-redshift galaxies 
detected in the second photometric catalog using {\sc BPZ}. 
For reliable color measurements, we PSF-match all the band images.
The photometric redshifts are used to both 
identify multiple images (see Section~\ref{sec:model}) and
select high-redshift galaxies (see Section~\ref{sec:dropout}).

\section{Mass Modeling Procedure}\label{sec:model}
Here we describe the method to model the mass distributions of the
four HFF clusters in detail. We adopt the so-called
  ``parametric lens modeling'' approach, in which a
  simply parametrized mass model consisting of several mass components
  is assumed and the model parameters are optimized to reproduce
  observed multiple image properties.
Throughout the paper mass modeling and analysis are performed using
the public software {\sc glafic} \citep{oguri10}, which has
extensively been used for strong lensing mass modeling of clusters
\citep[e.g.,][]{oguri12,oguri13,kohlinger14,ishigaki15,newman15}. 

\subsection{Mass Components}
In this paper we adopt the following mass components. Details of each
mass component are described in \citet{oguri10}. We give a brief
summary below.

A cluster-scale dark halo is modeled by an elliptical extension
of the NFW \citep{nfw97} density profile. We introduce an elliptical
symmetry in the projected mass density, and compute its lensing
properties by numerical integrals \citep{schramm90}. The model
parameters include virial mass $M$, positions, ellipticity
$e\equiv 1-a/b$ ($a$ and $b$ being minor and major axis lengths,
respectively) and its position angle $\theta_{e}$, and
concentration parameter $c$. 

Member galaxies are modeled by pseudo-Jaffe ellipsoids
\citep{keeton01}. To reduce the number of parameters, in most cases we
introduce scaling relations of model parameters with luminosity
$L$, such that velocity dispersion is given by
$\sigma / \sigma_{*} \propto L^{1/4}$ and truncation radius
$r_{\rm trun} / r_{\rm trun, *} \propto L^{\eta}$.
The ellipticity and position angle of each galaxy are fixed to the
values measured by {\sc SExtractor}.
All the input quantities for the member galaxies are measured 
in the \iFilter\ band. Luminosities are computed from total
magnitudes ({\tt MAG\_AUTO}) given by {\sc SExtractor}.
The model parameters are the normalization of velocity dispersion
$\sigma_{*}$, truncation radius $r_{\rm trun,*}$, and 
dimensionless parameter $\eta$.
We call this model of a set of member galaxies GALS.

Member galaxies that are located adjacent to multiple images can have
significant contributions to the image properties of the multiple
images including their locations. For some of these member galaxies we
do not apply the scaling relations mentioned above but instead model
them independently by pseudo-Jaffe ellipsoid components, to which we
refer as PJE. The model parameters are velocity dispersion $\sigma$,
ellipticity $e$ and its position angle $\theta_{e}$, and truncation
radius $r_{\rm trun}$. 

It has been shown that adding an external perturbation on
the lens potential and an internal perturbation describing
a possible asymmetry of the cluster mass distribution 
sometimes improves the mass model significantly
\citep[e.g.,][]{oguri10,oguri13}.
Both perturbations are described by a multipole Taylor
expansion at the position of the BCG of the form
$\phi = (C/m) r^{n}\cos m(\theta-\theta_{*})$, 
where $r$ is the distance from the BCG, $\theta$ is angular coordinate, 
$\theta_{*}$ is position angle, and $C$ is expansion coefficient.
In the case of the external perturbation, the zeroth ($n=0$, $m=0$) and 
the first ($n=1$, $m=1$) orders of the Taylor expansion are unobservable. 
We call the second order term of the external perturbation 
($n=2$, $m=2$), which is equivalent to
the so-called external shear, PERT. 
We also include higher multipole terms ($m\geq 3$) to 
approximately model higher-order terms of the external perturbation as well as 
a possible asymmetry of the cluster mass distribution, which we refer to as MPOLE.
Note that a term inducing constant convergence 
$\kappa$ ($n=2$, $m=0$) is not included in our mass modeling.

The amplitudes of the perturbations are defined for a given fiducial
source redshift $z_{s, {\rm fid}}$, and are scaled with the source
redshift assuming that the perturbation originates from the structure
at the cluster redshift. The model parameters for PERT are external
shear $\gamma$ and its position angle $\theta_{\gamma}$, and those for
MPOLE are expansion coefficient $\epsilon$, position angle
$\theta_{\epsilon}$, $m$, and $n$.
The values of $\gamma$ and $\epsilon$
are assumed to be constant over the entire field.

\subsection{Modeling Strategy}
We adopt the following unified strategy for conducting our mass
modeling. We place several NFW components on the positions of bright
cluster member galaxies. When an NFW component has a  
sufficient number of multiple images around it to constrain the model parameters
well, all the NFW model parameters are treated as free parameters. 
On the other hand, for NFW components located at the edge or outside
the strong lensing regions, we fix some model parameters such as positions,
ellipticities, and position angles, to observed values. 
For NFW components lacking strong observational constraints, it is also 
difficult to reliably constrain the concentration parameter $c$. In this
case we simply assume $c=10$.

We start with a small number of NFW components, 
and increase the number of components 
until we find the least reduced $\chi^{2}$.
We stop adding an NFW component when it begins to increase 
the reduced $\chi^{2}$, which is caused because a 
decrease in the degree of freedom 
surpasses an improvement in the raw $\chi^{2}$.
Perturbations (PERT and MPOLE) are also added as long as
they improve the mass model significantly. In parallel with
building the mass model, we iteratively refine multiple images used as
constraints, by validating known multiple image candidates and
searching for new multiple image candidates. New multiple image
candidates are identified based on consistency with the mass model
and on colors, morphologies, and photometric redshifts. 
Our selection of multiple images is conservative in
the sense that we remove any unreliable or suspicious candidates. 
A final set of multiple images for each cluster is given in
Section~\ref{sec:input}. 

About one fifth of the multiple images have spectroscopic redshifts.
The source redshifts are fixed to the spectroscopic redshifts when available.
The redshifts of the other multiple images are treated as model
parameters and are optimized together with source positions. 
Some multiple images have a precise photometric redshift
estimate. For them, we include this information in the optimization by adding a
Gaussian prior centered at the estimated redshift and a
conservative standard deviation of $\sigma_{z}=0.5$ (see also below).   
We choose this conservative value in order not to avoid any bias in
the best-fitting mass model originating from potential biases in our
photometric redshift estimates. 

\subsection{Optimizations and Error Estimates}
All the model parameters are simultaneously optimized to reproduce 
the positions and photometric redshifts of the multiple images.
Specifically, the optimization is performed to minimize $\chi^{2}$
\begin{eqnarray}
\chi^{2} &=& \chi^{2}_{\rm pos} + \chi^{2}_{z},\label{eq:chi2}\\
\chi^{2}_{\rm pos} &=& \sum_{i} \frac{\left|\mathbf{x}_{i, {\rm obs}}
  - \mathbf{x}_{i}\right |^{2}}{\sigma^{2}_{x_{i}}},\label{eq:chi2pos}\\
\chi^{2}_{z} &=& \sum_{j} \frac{(z_{j, {\rm obs}} - z_{j})^{2}}{\sigma^{2}_{z}},
\end{eqnarray}
where $\mathbf{x}_{i}$ is the position of the $i$-th image and $z_j$ is
the source redshift of the $j$-th system. The positional uncertainties
$\sigma_{x,i}$ can be different for different images and are given in
Section~\ref{sec:input}. For \clone, we include an
additional term $\chi^2_\mu=(\mu_{\rm obs}-\mu)^2/\sigma_\mu^2$ from
the observation of a Type Ia supernova behind this cluster (see
Section~\ref{sec:input} for more details).

Formally we need to solve a non-linear lens equation to estimate the
position $\chi^2$ (Equation~\ref{eq:chi2pos}), which is time-consuming.
We adopt the so-called source plane minimization which evaluates
Equation~(\ref{eq:chi2pos}) in the source plane. Once a distance in
the source plane is converted to a corresponding distance in the
image plane using the full magnification tensor, this provides a very
good approximation for the image plane position $\chi^2$ 
\citep[see Appendix 2 of][]{oguri10}. 

We derive the best-fitting mass model for each cluster that minimizes the
total $\chi^2$ (Equation~\ref{eq:chi2}) by a standard downhill simplex
method. In addition, we run Markov Chain Monte Carlo (MCMC) to
estimate errors in the mass models. When deriving the best-fitting 
mass model and running MCMC, the parameter
ranges of ellipticity, concentration parameter, and index
$\eta$ for GALS are restricted to [0, 0.8], [1, 40], and [0.2, 1.5],
respectively. 

\begin{figure*}[p]
  \centering
      \includegraphics[width=0.49\linewidth, trim=120 103 140 80, clip]{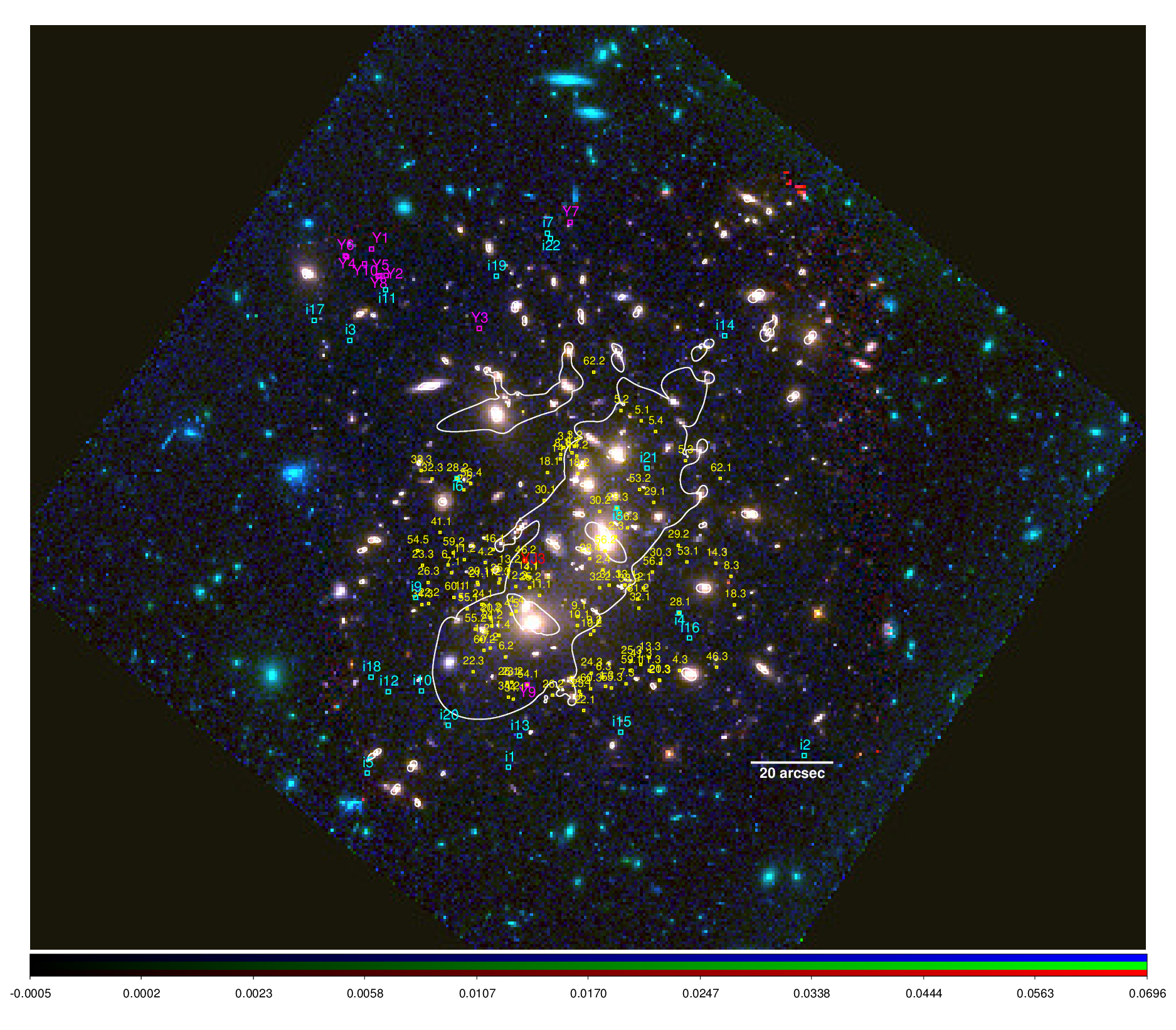}
      \includegraphics[width=0.49\linewidth, trim=40 40 130 50.5, clip]{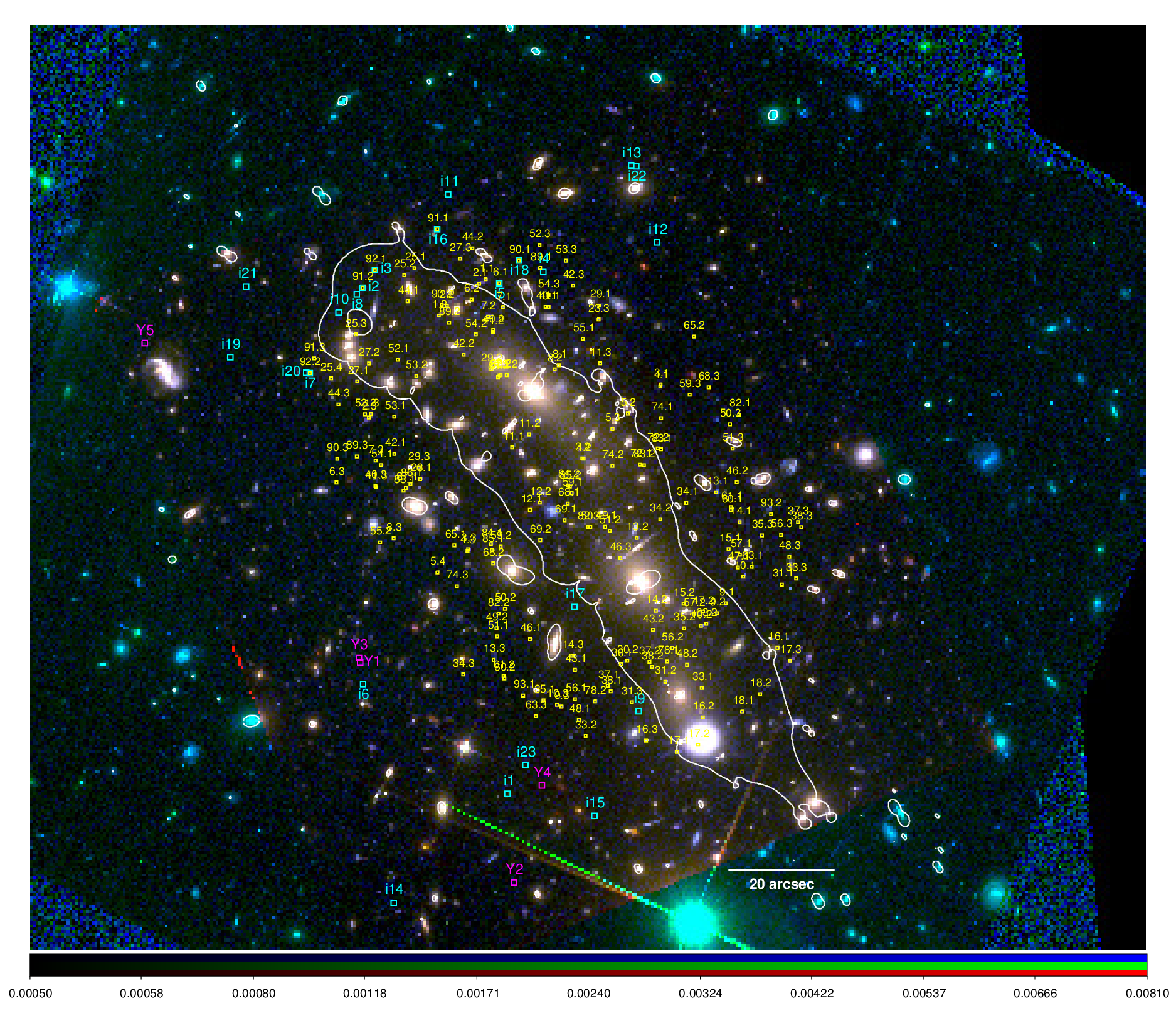}\\
      \vspace{2pt}
      \includegraphics[width=0.49\linewidth, trim=65 40 65 30, clip]{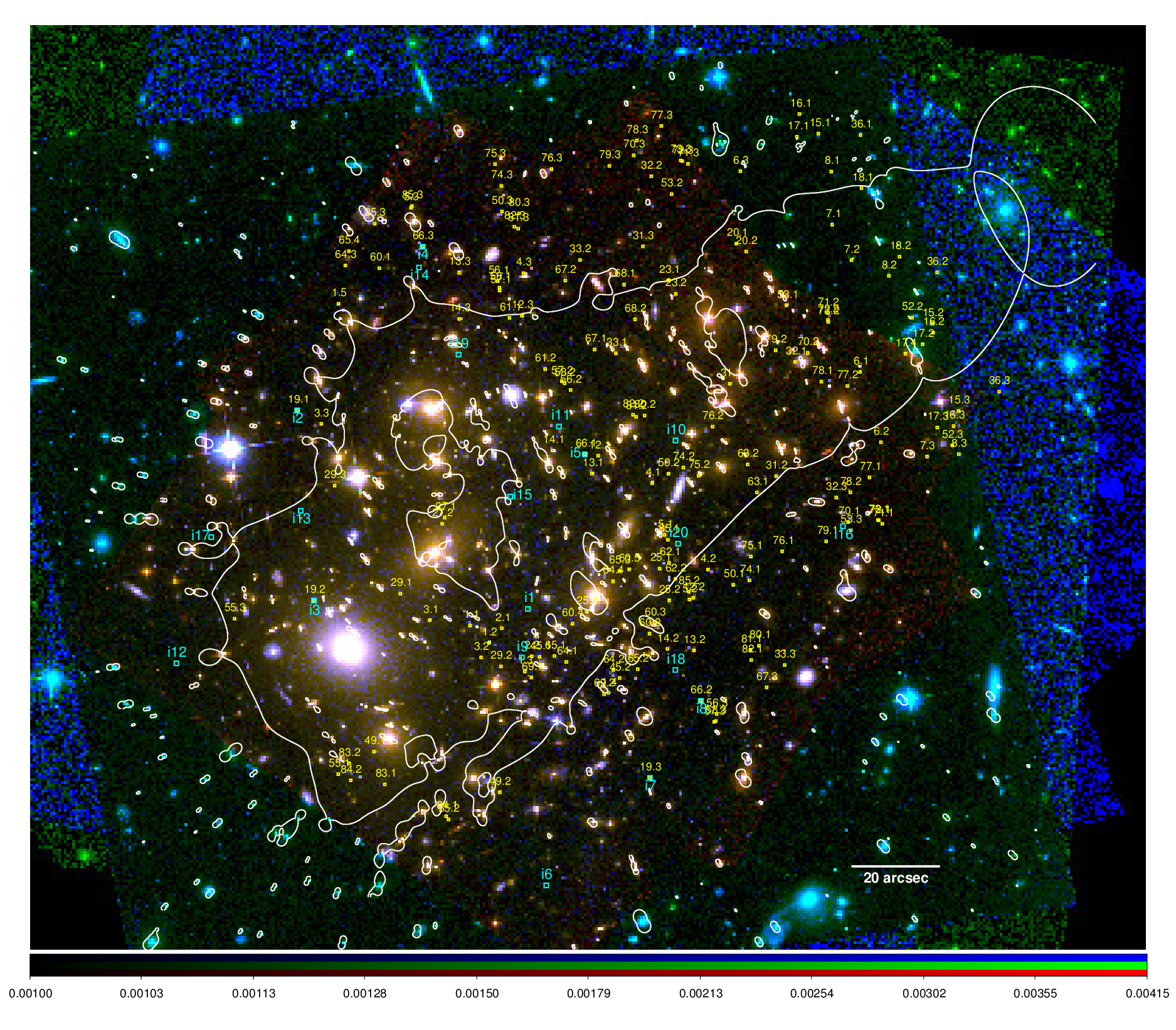}
      \includegraphics[width=0.49\linewidth, trim=66 85 110 30, clip]{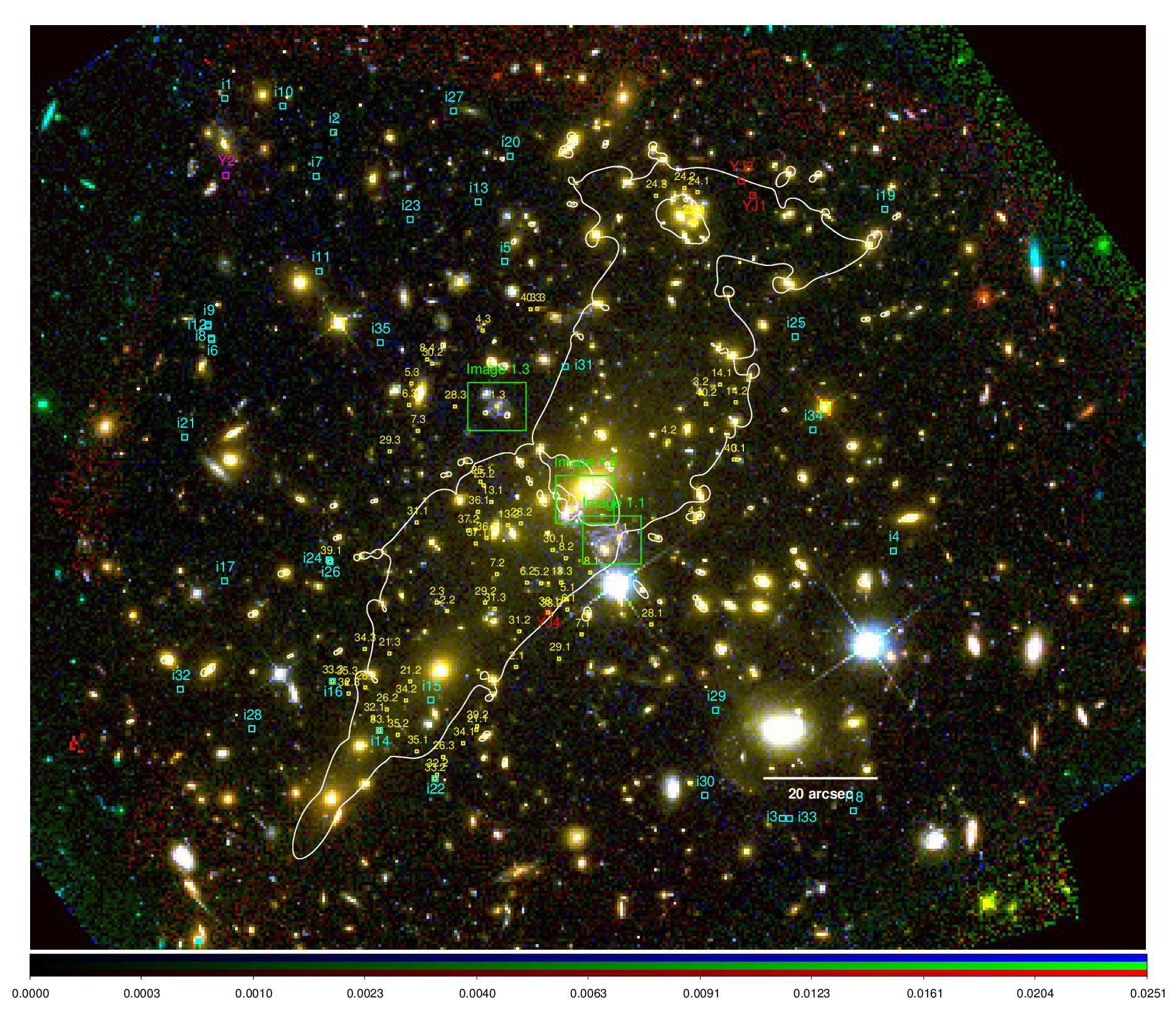}
      \includegraphics[width=0.325\linewidth, trim=15 40 15 0, clip]{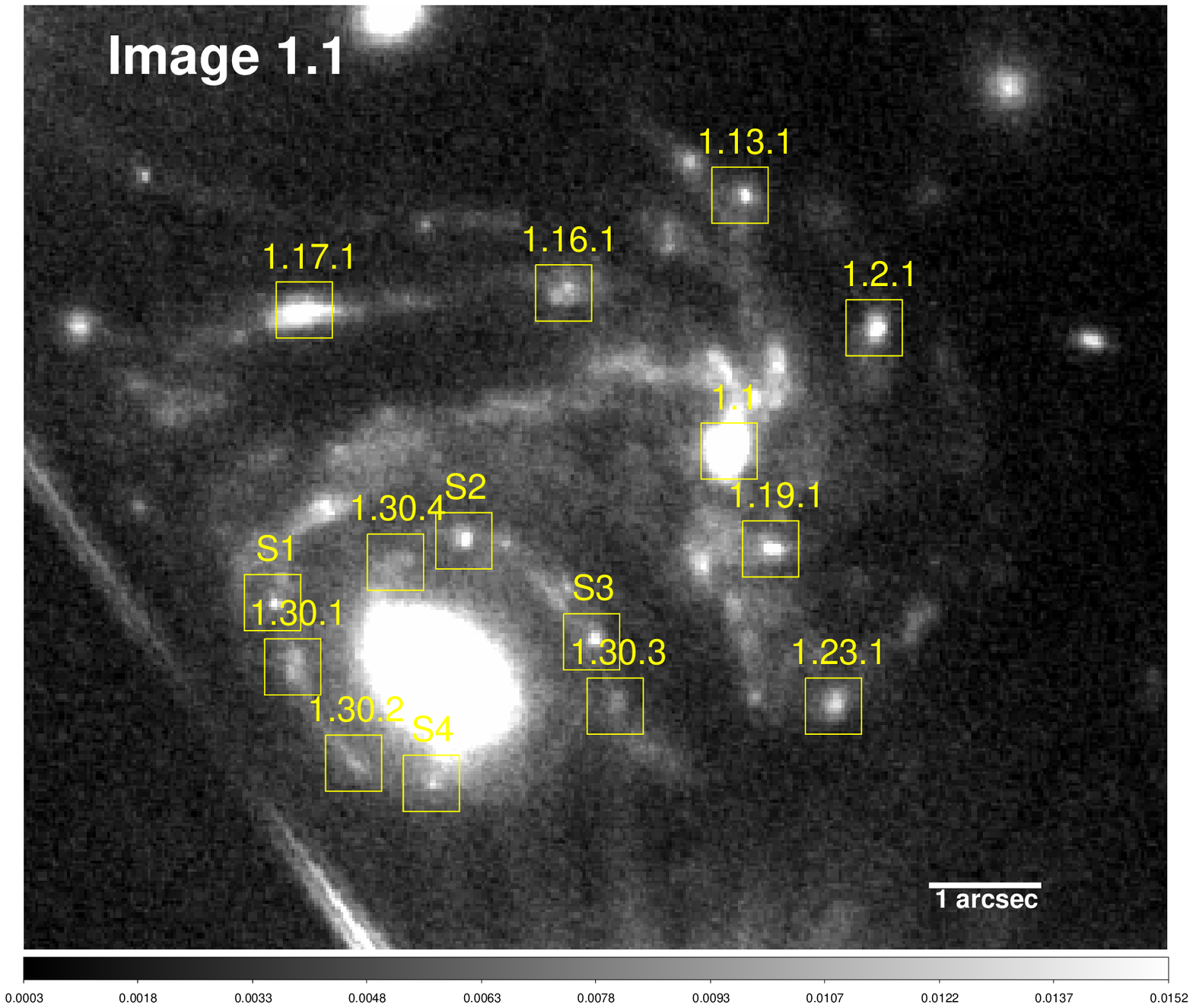}
      \includegraphics[width=0.325\linewidth, trim=15 40 15 0, clip]{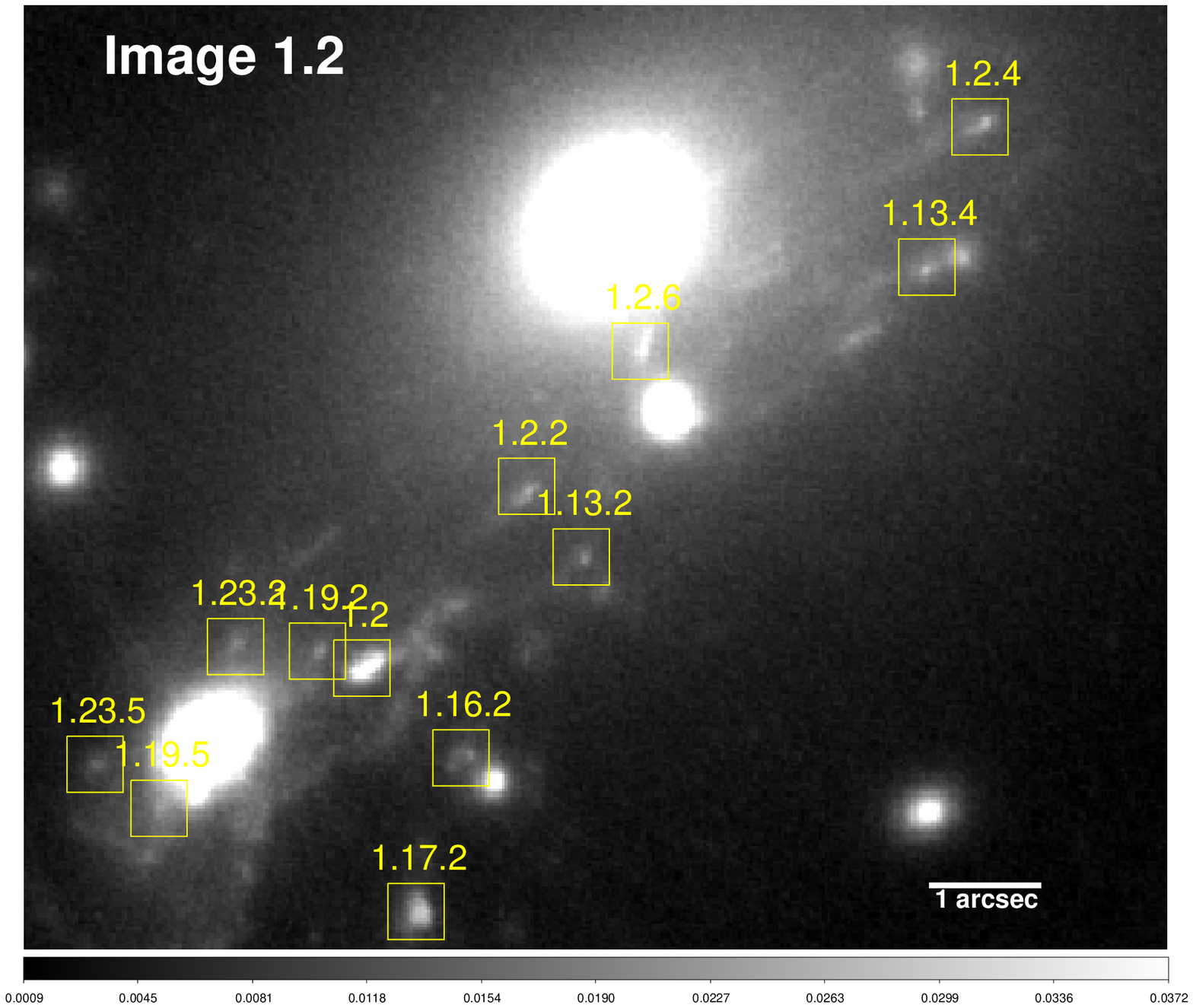}
      \includegraphics[width=0.325\linewidth, trim=15 40 15 0, clip]{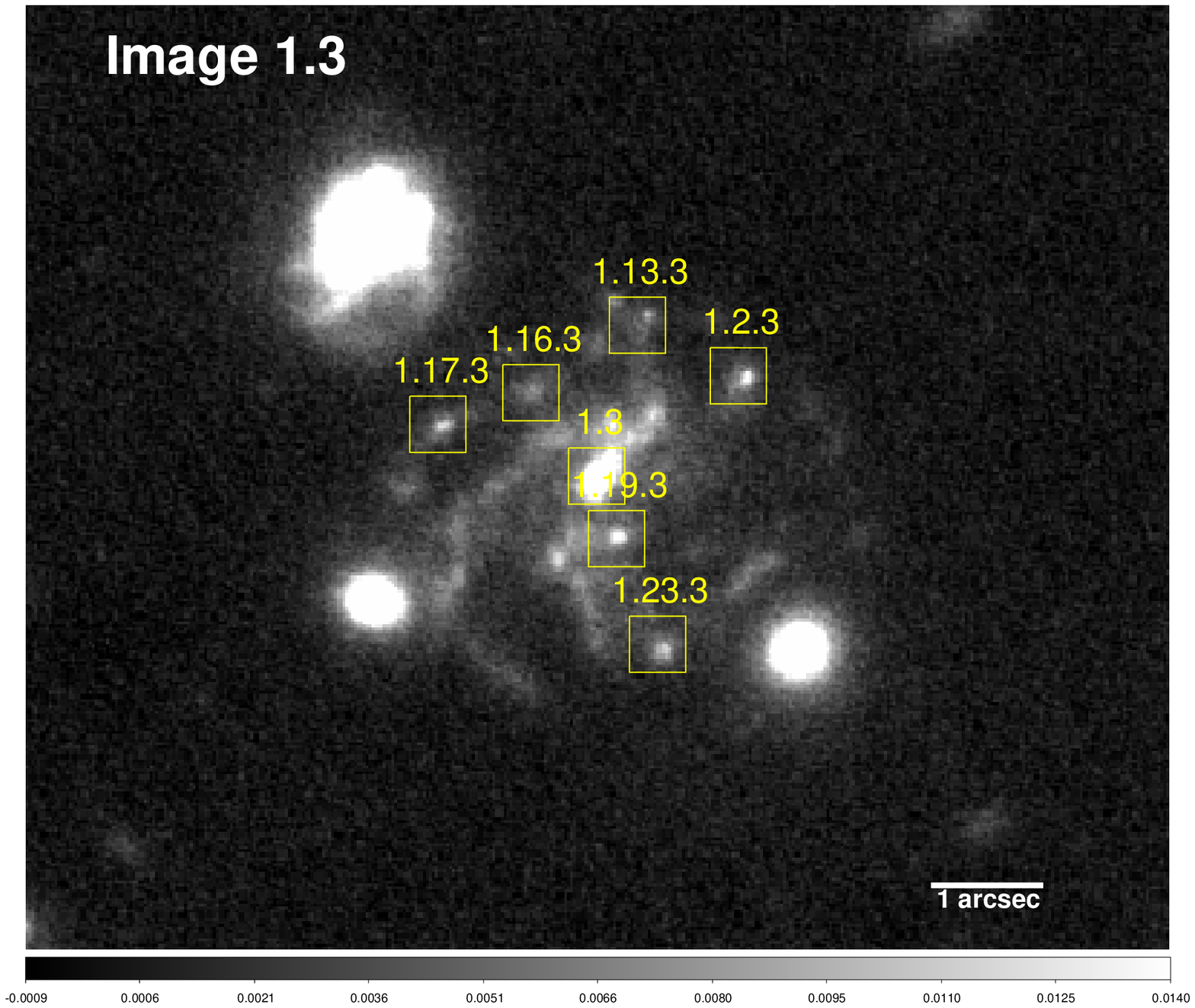}
  \caption{Multiple image systems used for mass modeling, dropout
  galaxies, and critical curves of the best-fitting models for \clone\ ({\it upper left}), 
  \cltwo\ ({\it upper right}), \clthree\ ({\it middle left}), and \clfour\ ({\it middle right}). 
  Underlying color-composite images are created from the {\it HST}
  \bFilter+\vFilter, \iFilter+\yFilter, \jFilter+\jhFilter+\hFilter\ band
  images. Small yellow squares show the positions of multiple images (see
  Appendix~\ref{sec:appendix_multiples} for the coordinates). High-redshift
  dropout galaxies are marked with large squares (see
  Section~\ref{sec:dropout} for details). Critical curves for a source redshift of $z=8$
   are shown with the solid lines. 
   Bottom panels show zoomed in {\it HST} \iFilter-band images of the 
   system 1 in the \clfour\ field. Small yellow squares represent the positions 
   of multiply imaged knots that are used as constraints in mass modeling.
   }
  \label{fig:colorimage}
\end{figure*}

\subsection{Input Data for Each Cluster}\label{sec:input}

\begin{deluxetable*}{lcccc}
\tabletypesize{\scriptsize}
\tablecaption{Summary of mass modeling\label{tab:modelsummaries}} 
\tablewidth{0pt}
\tablehead{
  \colhead{Cluster} & \colhead{\# of multiple image} & \colhead{\#
    of multiple images} & \colhead{$\chi^2$/dof} &  \colhead{Image
        plane RMS}\\
  \colhead{} &  \colhead{systems (with spec-$z$)} & \colhead{}
  & \colhead{} & \colhead{($''$)} 
}
\startdata
\clone   & $ 38\ (5)$  & $111$ & 98.2/100  & 0.37 \\
\cltwo   & $ 68\ (16)$ & $182$ & 155.8/168 & 0.44 \\
\clthree & $ 60\ (8)$  & $173$ & 144.5/144 & 0.52 \\
\clfour  & $ 36\ (16)$ & $108$ & 100.1/103 & 0.31 
\enddata
\end{deluxetable*}

\begin{deluxetable*}{cccccccc} 
\tabletypesize{\scriptsize}
\tablecaption{Mass Model Parameters for \clone\label{tab:modelparams_a2744}}
\tablewidth{0pt}
\tablehead{
      \colhead{Component} & \colhead{Model} & \colhead{Mass} & \colhead{$e$} & \colhead{$\theta_{e}$} & \colhead{$c$} & \colhead{$\Delta x$\tablenotemark{a}} & \colhead{$\Delta y$\tablenotemark{a}}\\
      \colhead{} & \colhead{} & \colhead{$(10^{14}\ h^{-1}M_{\odot})$} & \colhead{} & \colhead{(deg)} & \colhead{} & \colhead{(arcsec)} & \colhead{(arcsec)}
}
\startdata
Cluster halo 1 & NFW & $4.46^{+2.35}_{-1.36}$ & $0.45^{+0.04}_{-0.04}$ & $152.23^{+2.26}_{-2.56}$ & $3.39^{+0.66}_{-0.53}$ & $-0.28^{+0.37}_{-0.40}$ & $0.20^{+0.42}_{-0.43}$ \\ 
Cluster halo 2 & NFW & $1.57^{+0.23}_{-0.24}$ & $0.42^{+0.03}_{-0.03}$ & $132.78^{+1.67}_{-0.00}$ & $8.46^{+0.54}_{-0.49}$ & $-18.51^{+0.26}_{-0.26}$ & $-17.91^{+0.22}_{-0.23}$ \\ 
Cluster halo 3 & NFW & $0.36^{+0.07}_{-0.06}$ & $0.79^{+0.01}_{-0.02}$ & $101.97^{+2.87}_{-2.81}$ & [$10.00$] & [$-26.97$] & [$30.91$] \\ 
\hline
 &  & $\sigma_{*}\tablenotemark{b}$ & $r_{\rm trun, *}$ & $\eta$ & & \\
 &  & $({\rm km\ s^{-1}})$ & $('')$ & & & \\ \hline \\
Member galaxies  & GALS & $208.73^{+7.61}_{-7.77}$ & $82.71^{+29.72}_{-28.75}$ & $1.25^{+0.16}_{-0.20}$ & \\ 
\hline 
 &  & $z_{s, {\rm fid}}$ & $\gamma$ & $\theta_{\gamma}$ & & \\
 &  & & & (deg) & & & \\\hline \\
External perturbation & PERT & [$2.00$] & $0.05^{+0.02}_{-0.02}$ & $138.75^{+4.24}_{-7.63}$ & \\ 
\hline
 &  & $z_{s, {\rm fid}}$ & $ \epsilon $ & $\theta_{\rm \epsilon}$ & $m$ & $n$ & \\
& & & & (deg) & &  & \\\hline \\
Multipole perturbation & MPOLE & [$2.00$] & $0.01^{+0.00}_{-0.00}$ & $85.41^{+3.31}_{-2.88}$ & [$3.00$] & [$2.00$]
\enddata
\tablenotetext{a}{Coordinates are relative to the brightest cluster galaxy position in the \clone\ field (R.A. = 3.58611, Decl. = $-30.40024$).}
\tablenotetext{b}{The normalization luminosity $L^*$ corresponds to $i_{814} = 18.33$.}
\tablenotetext{*}{Numbers in square brackets are fixed during the model optimization.}
\end{deluxetable*}

\begin{deluxetable*}{cccccccc} 
\tabletypesize{\scriptsize}
\tablecaption{Mass Model Parameters for \cltwo\label{tab:modelparams_m0416}}
\tablewidth{0pt}
\tablehead{
      \colhead{Component} & \colhead{Model} & \colhead{Mass} & \colhead{$e$} & \colhead{$\theta_{e}$} & \colhead{$c$} & \colhead{$\Delta x$\tablenotemark{a}} & \colhead{$\Delta y$\tablenotemark{a}}\\
      \colhead{} & \colhead{} & \colhead{$(10^{14}\ h^{-1}M_{\odot})$} & \colhead{} & \colhead{(deg)} & \colhead{} & \colhead{(arcsec)} & \colhead{(arcsec)}
}
\startdata
Cluster halo 1 & NFW & $3.22^{+1.06}_{-0.76}$ & $0.64^{+0.02}_{-0.02}$ & $56.59^{+1.69}_{-1.64}$ & $4.20^{+0.42}_{-0.37}$ & $-2.09^{+0.65}_{-0.51}$ & $1.39^{+0.42}_{-0.44}$ \\ 
Cluster halo 2 & NFW & $1.90^{+0.97}_{-0.66}$ & $0.64^{+0.03}_{-0.03}$ & $42.07^{+1.70}_{-1.51}$ & $5.31^{+1.26}_{-0.94}$ & $22.18^{+0.76}_{-0.67}$ & $-35.65^{+0.51}_{-0.57}$ \\ 
Cluster halo 3 & NFW & $0.51^{+0.21}_{-0.15}$ & $0.61^{+0.08}_{-0.08}$ & $40.41^{+3.86}_{-4.17}$ & $5.95^{+1.30}_{-1.02}$ & $25.68^{+0.97}_{-1.38}$ & $-55.26^{+1.62}_{-1.02}$ \\ 
\hline
 &  & $\sigma_{*}\tablenotemark{b}$ & $r_{\rm trun, *}$ & $\eta$ & & \\
 &  & $({\rm km\ s^{-1}})$ & $('')$ & & & \\ \hline \\
Member galaxies  & GALS & $261.59^{+21.36}_{-17.44}$ & $24.04^{+10.02}_{-6.09}$ & $1.31^{+0.12}_{-0.15}$ & \\ 
\hline
 &  & $\sigma$ & $e$ & $\theta_{e}$ & $r_{\rm trun}$ & \colhead{$\Delta x$\tablenotemark{a}} & \colhead{$\Delta y$\tablenotemark{a}} \\
 &  & $({\rm km\ s^{-1}})$ & & (deg) & $('')$ & \colhead{(arcsec)} & \colhead{(arcsec)} \\ \hline \\
Member galaxy  & PJE & $125.29^{+50.22}_{-19.36}$ & [$0.27$] & [$166.70$] & $0.70^{+1.38}_{-0.47}$ & [$-14.56$] & [$15.28$] \\ 
\hline
 &  & $z_{s, {\rm fid}}$ & $\gamma$ & $\theta_{\gamma}$ &  & \\
 &  & & & (deg) & & & \\\hline \\
External perturbation & PERT & [$2.00$] & $0.06^{+0.01}_{-0.01}$ & $35.50^{+3.83}_{-4.40}$ \\ 
\hline
 &  & $z_{s, {\rm fid}}$ & $ \epsilon $ & $\theta_{\rm \epsilon}$ & $m$ & $n$ & \\
 &  & & & (deg) & & & \\\hline \\
Multipole perturbation & MPOLE & [$2.00$] & $0.01^{+0.00}_{-0.00}$ & $65.07^{+5.96}_{-5.10}$ & [$3.00$] & [$2.00$] 
\enddata
\tablenotetext{a}{Coordinates are relative to the brightest cluster galaxy position in the \cltwo\ field (R.A. = 64.0380981, Decl. = $-24.0674834$).}
\tablenotetext{b}{The normalization luminosity $L^*$ corresponds to $i_{814} = 18.73$.}
\tablenotetext{*}{Numbers in square brackets are fixed during the model optimization.}
\end{deluxetable*}

\begin{deluxetable*}{ccccccccc} 
\tabletypesize{\scriptsize}
\tablecaption{Mass Model Parameters for \clthree\label{tab:modelparams_m0717}}
\tablewidth{0pt}
\tablehead{
      \colhead{Component} & \colhead{Model} & \colhead{Mass} & \colhead{$e$} & \colhead{$\theta_{e}$} & \colhead{$c$} & \colhead{$\Delta x$\tablenotemark{a}} & \colhead{$\Delta y$\tablenotemark{a}} &\\
      \colhead{} & \colhead{} & \colhead{$(10^{14}\ h^{-1}M_{\odot})$} & \colhead{} & \colhead{(deg)} & \colhead{} & \colhead{(arcsec)} & \colhead{(arcsec)} &
}
\startdata
Cluster halo 1 & NFW & $4.78^{+1.02}_{-0.59}$ & $0.63^{+0.03}_{-0.03}$ & $135.70^{+1.43}_{-1.09}$ & $3.97^{+0.36}_{-0.53}$ & $8.09^{+1.05}_{-0.62}$ & $3.44^{+1.11}_{-0.79}$ \\ 
Cluster halo 2 & NFW & $2.02^{+0.20}_{-0.39}$ & $0.73^{+0.02}_{-0.02}$ & $135.60^{+0.95}_{-1.46}$ & $3.96^{+0.23}_{-0.31}$ & $35.81^{+1.13}_{-0.79}$ & $-9.95^{+0.97}_{-0.89}$ \\ 
Cluster halo 3 & NFW & $2.23^{+0.25}_{-0.29}$ & $0.56^{+0.02}_{-0.02}$ & $142.09^{+0.97}_{-1.32}$ & $6.98^{+1.02}_{-1.49}$ & $-2.12^{+0.52}_{-0.48}$ & $30.13^{+1.11}_{-0.79}$ & \\ 
Cluster halo 4 & NFW & $3.18^{+0.48}_{-0.28}$ & $0.31^{+0.04}_{-0.04}$ & $152.04^{+2.86}_{-1.97}$ & $3.85^{+0.31}_{-0.39}$ & $67.18^{+0.58}_{-0.61}$ & $49.61^{+0.61}_{-0.57}$ & \\ 
Cluster halo 5 & NFW & $1.51^{+0.08}_{-0.13}$ & [$0.32$] & [$174.30$] & [$10.00$] & [$129.13$] & [$77.20$] & \\ 
Cluster halo 6 & NFW & $0.56^{+0.13}_{-0.13}$ & $0.19^{+0.23}_{-0.12}$ & $105.68^{+1.57}_{-2.84}$ & $2.39^{+0.53}_{-0.55}$ & [$-19.33$] & [$-21.66$] & \\ 
Cluster halo 7 & NFW & $1.20^{+0.16}_{-0.19}$ & $0.55^{+0.04}_{-0.04}$ & $129.81^{+1.33}_{-1.61}$ & $3.56^{+0.49}_{-0.53}$ & [$108.64$] & [$45.46$] & \\ 
Cluster halo 8 & NFW & $0.14^{+0.02}_{-0.03}$ & $0.78^{+0.02}_{-0.02}$ & $146.46^{+4.24}_{-1.51}$ & $2.69^{+0.56}_{-0.68}$ & [$-10.32$] & [$-42.04$] & \\ 
Cluster halo 9 & NFW & $0.06^{+0.05}_{-0.02}$ & $0.76^{+0.03}_{-0.04}$ & $133.99^{+4.15}_{-4.48}$ & $12.40^{+5.93}_{-4.83}$ & $29.63^{+2.05}_{-2.53}$ & $-32.35^{+1.20}_{-0.99}$ & \\ 
\hline
 &  & $\sigma_{*}\tablenotemark{b}$ & $r_{\rm trun, *}$ & $\eta$ & & \\
 &  & $({\rm km\ s^{-1}})$ & $('')$ & & & \\ \hline \\
Member galaxies  & GALS & $518.64^{+35.12}_{-43.24}$ & $7.06^{+2.25}_{-2.08}$ & $0.43^{+0.07}_{-0.09}$ & \\ 
\hline
 &  & $z_{s, {\rm fid}}$ & $\gamma$ & $\theta_{\gamma}$ & & & \\
 &  & & & (deg) & & & & \\\hline \\
External perturbation & PERT & [$2.00$] & $0.12^{+0.00}_{-0.01}$ & $51.13^{+1.49}_{-0.90}$ & \\ 
\hline
 &  & $z_{s, {\rm fid}}$ & $ \epsilon $ & $\theta_{\rm \epsilon}$ & $m$ & $n$ & \\
 &  & & & (deg) & & & & \\\hline \\
Multipole perturbation 1 & MPOLE & [$2.00$] & $0.02^{+0.00}_{-0.00}$ & $42.99^{+2.36}_{-7.54}$ & [$3.00$] & [$2.00$]\\ 
Multipole perturbation 2 & MPOLE & [$2.00$] & $0.01^{+0.00}_{-0.00}$ & $8.54^{+2.99}_{-1.74}$ & [$4.00$] & [$2.00$]\\ 
Multipole perturbation 3 & MPOLE & [$2.00$] & $0.01^{+0.00}_{-0.00}$ & $20.00^{+1.44}_{-1.75}$ & [$5.00$] & [$2.00$]
\enddata
\tablenotetext{a}{Coordinates are relative to the brightest cluster galaxy position in the \clthree\ field (R.A. = 109.3982391, Decl. = $+37.7457307$).}
\tablenotetext{b}{The normalization luminosity $L^*$ corresponds to $i_{814} = 17.16$.}
\tablenotetext{*}{Numbers in square brackets are fixed during the model optimization.}
\end{deluxetable*}

\begin{deluxetable*}{cccccccc} 
\tabletypesize{\scriptsize}
\tablecaption{Mass Model Parameters for \clfour\label{tab:modelparams_m1149}}
\tablewidth{0pt}
\tablehead{
      \colhead{Component} & \colhead{Model} & \colhead{Mass} & \colhead{$e$} & \colhead{$\theta_{e}$} & \colhead{$c$} & \colhead{$\Delta x$\tablenotemark{a}} & \colhead{$\Delta y$\tablenotemark{a}}\\
      \colhead{} & \colhead{} & \colhead{$(10^{14}\ h^{-1}M_{\odot})$} & \colhead{} & \colhead{(deg)} & \colhead{} & \colhead{(arcsec)} & \colhead{(arcsec)}
}
\startdata
Cluster halo 1 & NFW & $8.26^{+1.52}_{-1.83}$ & $0.49^{+0.02}_{-0.02}$ & $126.37^{+1.31}_{-1.28}$ & $3.82^{+0.34}_{-0.25}$ & $-0.21^{+0.15}_{-0.16}$ & $-0.12^{+0.10}_{-0.10}$ \\ 
Cluster halo 2 & NFW & $1.61^{+0.59}_{-0.46}$ & $0.67^{+0.09}_{-0.14}$ & $76.36^{+7.39}_{-6.88}$ & $6.66^{+2.75}_{-2.16}$ & [$16.38$] & [$47.36$] \\ 
Cluster halo 3 & NFW & $0.64^{+0.75}_{-0.34}$ & $0.70^{+0.05}_{-0.06}$ & $158.13^{+3.19}_{-4.23}$ & $2.57^{+1.50}_{-0.85}$ & $-22.93^{+1.04}_{-0.68}$ & $-32.21^{+1.31}_{-1.26}$ \\ 
Cluster halo 4 & NFW & $0.16^{+0.04}_{-0.03}$ & $0.68^{+0.08}_{-0.10}$ & $150.23^{+2.05}_{-2.71}$ & [$10.00$] & [$-44.77$] & [$-54.86$] \\ 
\hline
 &  & $\sigma_{*}\tablenotemark{b}$ & $r_{\rm trun, *}$ & $\eta$ & & \\
 &  & $({\rm km\ s^{-1}})$ & $('')$ & & & \\ \hline \\
Member galaxies  & GALS & $233.07^{+21.42}_{-16.80}$ & $2.88^{+1.07}_{-0.65}$ & $0.26^{+0.08}_{-0.04}$ & \\ 
\hline
 &  & $\sigma$ & $e$ & $\theta_{e}$ & $r_{\rm trun}$ &  \colhead{$\Delta x$\tablenotemark{a}} & \colhead{$\Delta y$\tablenotemark{a}} \\
 &  & $({\rm km\ s^{-1}})$ & & (deg) & $('')$ & \colhead{(arcsec)} & \colhead{(arcsec)} \\ \hline \\
Member galaxy\tablenotemark{c}  & PJE & $232.08^{+25.96}_{-17.38}$ & [$0.30$] & [$47.50$] & $1.26^{+0.54}_{-0.38}$ & [$3.22$] & [$-11.11$] \\ 
\hline
 &  & $z_{s, {\rm fid}}$ & $\gamma$ & $\theta_{\gamma}$ & & & \\
 &  & & & (deg) &  & & \\\hline \\
External perturbation & PERT & [$2.00$] & $0.04^{+0.02}_{-0.01}$ & $82.76^{+9.28}_{-7.17}$ & \\ 
\hline
 &  & $z_{s, {\rm fid}}$ & $ \epsilon $ & $\theta_{\rm \epsilon}$ & $m$ & $n$ \\
 &  & & & (deg) & &  & \\\hline \\
Multipole perturbation & MPOLE & [$2.00$] & $0.02^{+0.00}_{-0.00}$ & $165.09^{+4.04}_{-3.37}$ & [$3.00$] & [$2.00$] 
\enddata
\tablenotetext{a}{Coordinates are relative to the brightest cluster galaxy position in the \clfour\ field (R.A. = 177.3987491, Decl. = $+22.3985308$).}
\tablenotetext{b}{The normalization luminosity $L^*$ corresponds to $i_{814} = 18.80$.}
\tablenotetext{c}{This component corresponds to the member galaxy that produces four multiple images S1--S4 of SN Refsdal.}
\tablenotetext{*}{Numbers in square brackets are fixed during the model optimization.}
\end{deluxetable*}

\subsubsection{\clone}
Multiple images for this cluster have been identified in 
\citet{merten11}, \citet{atek14}, \citet{richard14}, \citet{zitrin14}, \citet{lam14}, 
\citet{ishigaki15}, and \citet{jauzac15}.
Spectroscopic redshifts of multiple images have been presented in
\citet{richard14}, \citet{johnson14}, and \citet{wang15}. 
\citet{lam14} and \citet{wang15} regarded systems 55 and 56 as
a part of systems 1 and 2, respectively, and assigned 
their redshifts accordingly.
To avoid introducing biases, we do not fix the redshifts but treat
them as model parameters. 
While \citet{wang15} reported the redshift of system 56 to be $z=1.2$
with a rating of probable, \citet{johnson14} estimated it to be 
$z=2.2$ and \citet{lam14} adopted this value in their mass modeling.
In our mass modeling, we do not assume any spectroscopic redshift 
on this system and find a model-predicted redshift of
$z=1.87^{+0.07}_{-0.07}$, which is closer to that of \citet{johnson14}.
Due to a controversy over the position of the counter image
 of system 3 \citep[see e.g.][for more details]{lam14, jauzac15}, we do not 
 use its position as a constraint in our mass modeling.
For system 5, we find one new counter image.
Although \citet{wang15} recently reported the redshift of system
22 to be $z=4.84$, we do not adopt this value because it
is not very secure. We identify a new set of multiple images (system
62) in the northwest part of this cluster.
As noted above, we conservatively exclude some multiple images in the literature.
As a result, we have 37 multiple image systems from the literature
and one new system for our mass modeling. The total
number of multiple images is 111. The positional uncertainty of
$\sigma_x=0\farcs4$ is assumed for all of them.  

In addition we include a magnification constraint at the position of
the type Ia supernova HFF14Tom at $z=1.3457$ \citep{rodney15}. 
The magnification of the HFF14Tom is estimated by a careful
cosmology-independent analysis to be $\mu=2.03\pm0.29$. We use this
constraint by adding a term to the total $\chi^2$
(Equation~\ref{eq:chi2}). 

\subsubsection{\cltwo}
Multiple images for this cluster have been identified in
\citet{zitrin13}, \citet{jauzac14}, and \citet{diego15a}. 
Spectroscopic redshifts of multiple images have been presented
in \citet{christensen12} and \citet{grillo15}.
We also use new spectroscopic redshifts from GLASS (Hoag et al. in
prep.; see also \citealt{schmidt14} and \citealt{treu15a}) and Rodney
et al. (in prep.). 
While \citet{jauzac14} estimated the redshift of system 14 to be
$z=2.0531$, \citet{grillo15} reported that its correct redshift is
$z=1.637$. We adopt the latter as it reproduces its image positions
well. We correct the positions of five counter images, 29.2,
37.3, 40.3, 41.3, and 55.2, and add nine new systems, 74, 78,
82, 83, 89, 90, 91, 92, and 93, and identify four new counter images,
6.3, 8.3, 34.3, and 50.3. 
As a result, we have 59 multiple image systems from the literature 
and nine new systems for our mass modeling. 
The total number of multiple images is 182.
The positional uncertainty of $\sigma_x=0\farcs4$ is assumed for all
of them.  

\subsubsection{\clthree}
Multiple images for this cluster have been identified in
\citet{zitrin09b}, \citet{limousin12}, \citet{vanzella14}, \citet{richard14}, and \citet{diego15b}. 
Spectroscopic redshifts of multiple images have been presented
in \citet{limousin12}, \citet{schmidt14}, \citet{vanzella14}, and \citet{treu15a}.
The redshift of system 5 was newly confirmed and those of systems 12 and 13
were updated by GLASS \citep{schmidt14, treu15a}.
While we use the updated redshift of system 12, we do not use 
that of system 5 as it is significantly different from our model prediction
and that of system 13 as it is less precise than that estimated 
in \citet{limousin12}.
We assign image 25.4 to system 25, which was regarded as a part of 
system 5 in \citet{diego15b}. 
We add six new counter images, 25.4, 55.3, 64.3, 64.4,
65.3, and 65.4, and 20 new systems, $66-85$.
As a result, we have 40 multiple image systems from the literature 
and 20 new systems for our mass modeling. The total number of
multiple images is 173.
As a foreground galaxy located at $({\rm R.A.} = 109.405027, 
{\rm Decl.} = +37.739714)$ makes a significant contribution to
the lensing effect, we independently model this galaxy 
by an NFW, but at the cluster redshift (Cluster halo 6) 
because {\sc glafic} does not support multiple lens planes.
We assume a positional uncertainty of $\sigma_x=0\farcs6$,
which is larger than those for the other HFF clusters, for all
multiple images.
The larger positional uncertainty and the large number of mass components
are due to the fact that the mass distribution of this cluster appears to be
considerably more complicated than the other clusters, presumably
due to ongoing multiple mergers \citep[see, e.g.,][]{limousin12}. 

\subsubsection{\clfour}
Multiple images for this cluster have been identified in
\citet{zitrin09a}, \citet{smith09}, \cite{zheng12}, \citet{rau14}, \citet{richard14}, 
\citet{jauzac16}, and \citet{treu15b}. Spectroscopic redshifts of multiple images
have been presented in \citet{smith09}, \citet{jauzac16}, \citet{grillo16}, and 
Brammer et al. (in prep.).  
While \citet{smith09} estimated the redshift of system 3 to be 
$z=2.497$, a recent study using GMOS and MUSE data \citep{jauzac16,grillo16}
revised its redshift to be $z=3.129$, which we adopt in our analysis. 
The new spectroscopic redshifts of system 13 by 
GLASS (Brammer et al. in prep.) and systems 4, 14, and 29 by MUSE  \citep{grillo16}
are used in our mass modeling \citep[see also][]{treu15b}.
 As a result, we have 10 multiple image systems from
the literature and 18 new systems, $21-40$, 
for our mass modeling.
We also include additional positional constraints from multiple images
of seven knots in a lensed face-on spiral galaxy at $z=1.488$ as well as
four supernova images of SN Refsdal in the lensed spiral galaxy
\citep{kelly15}.  The total number of multiple images is 108 from 36
systems.

In order to accurately predict the reappearance of SN Refsdal image
\citep{oguri15,sharon15,diego15c,jauzac16,grillo16} and its magnification, we
follow \citet{oguri15} to adopt different 
positional errors for different multiple images. Specifically, we
assume the standard positional error of $\sigma_x=0\farcs4$ for most
multiple images, but use a smaller error of $\sigma_x=0\farcs2$ for the
core and knots of the lensed spiral galaxy, and an even smaller error
of $\sigma_x=0\farcs05$ for the four SN images. 
A member galaxy located at R.A. $= 177.397784$, 
Decl. $= +22.395446$ clearly has a significant impact on
the prediction of the quadruple images S1--S4 of SN Refsdal. 
Thus we model this galaxy separately by a PJE.

\section{Mass Modeling Results}\label{sec:result}

\begin{figure*}[t]
  \centering
      \includegraphics[width=0.49\linewidth, trim=9 10 9 10, clip]{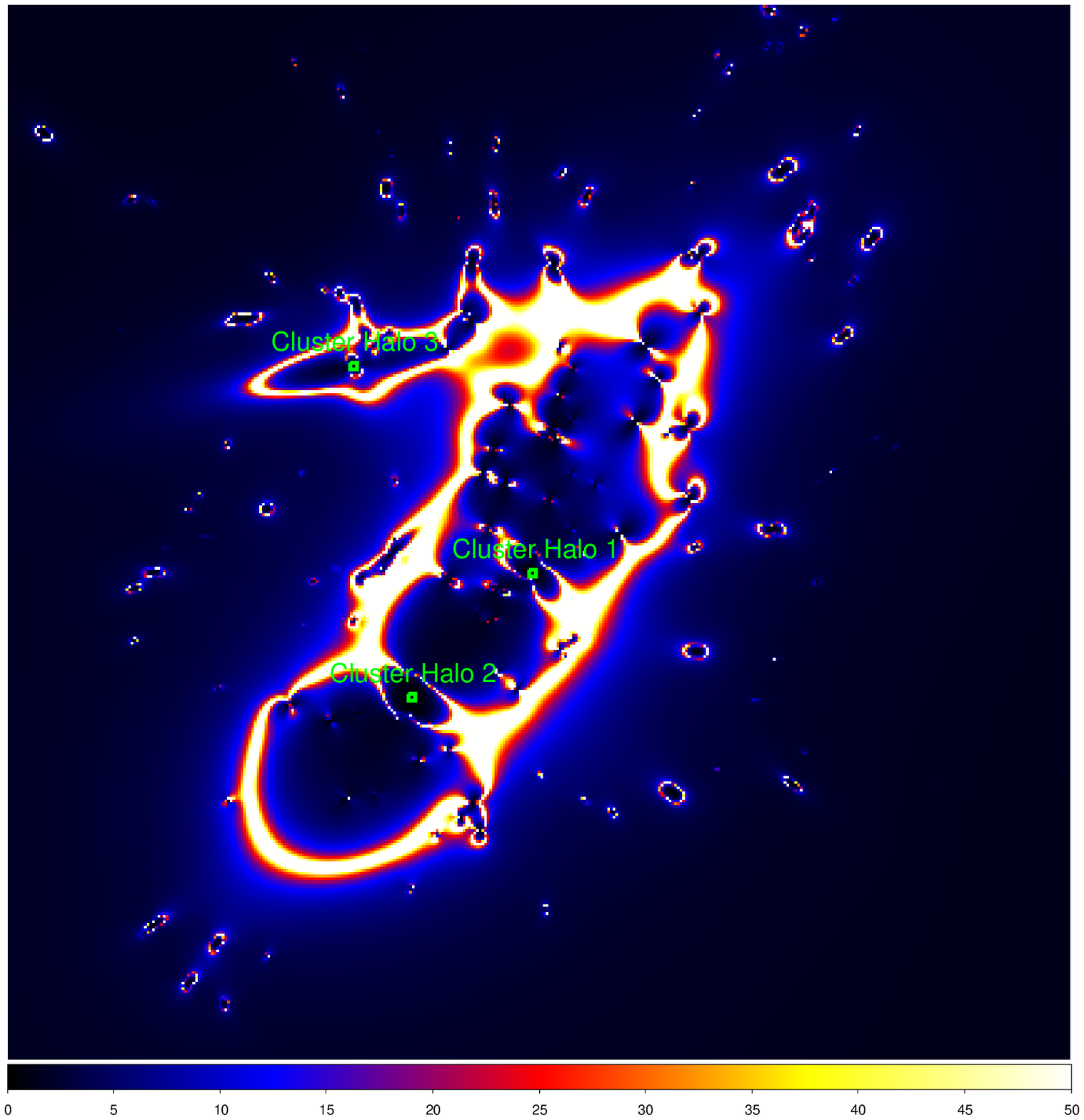}
      \includegraphics[width=0.49\linewidth, trim=9 10 9 10, clip]{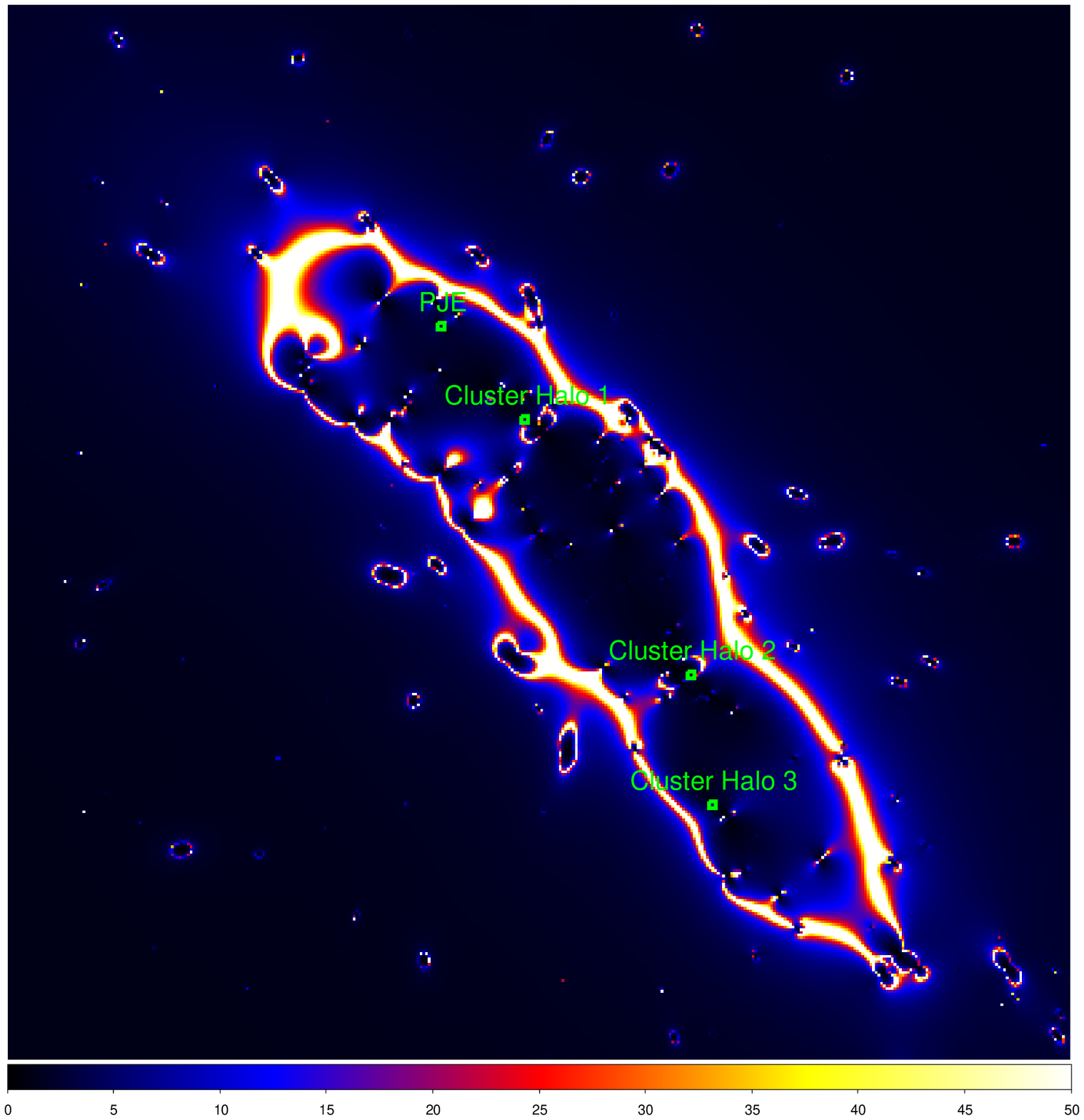}\\
      \vspace{2pt}
      \includegraphics[width=0.49\linewidth, trim=9 10 9 10, clip]{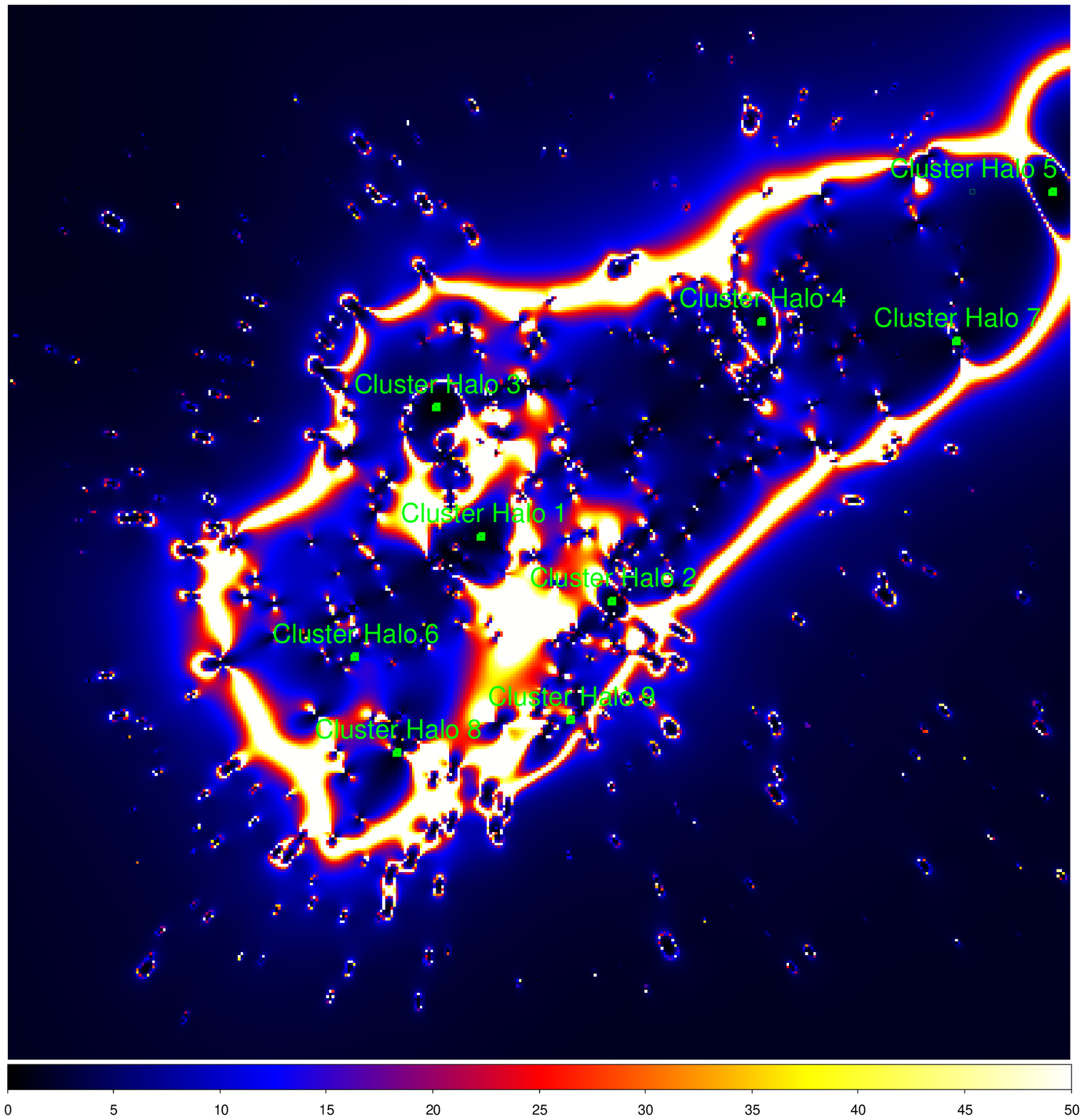}
      \includegraphics[width=0.49\linewidth, trim=9 10 9 10, clip]{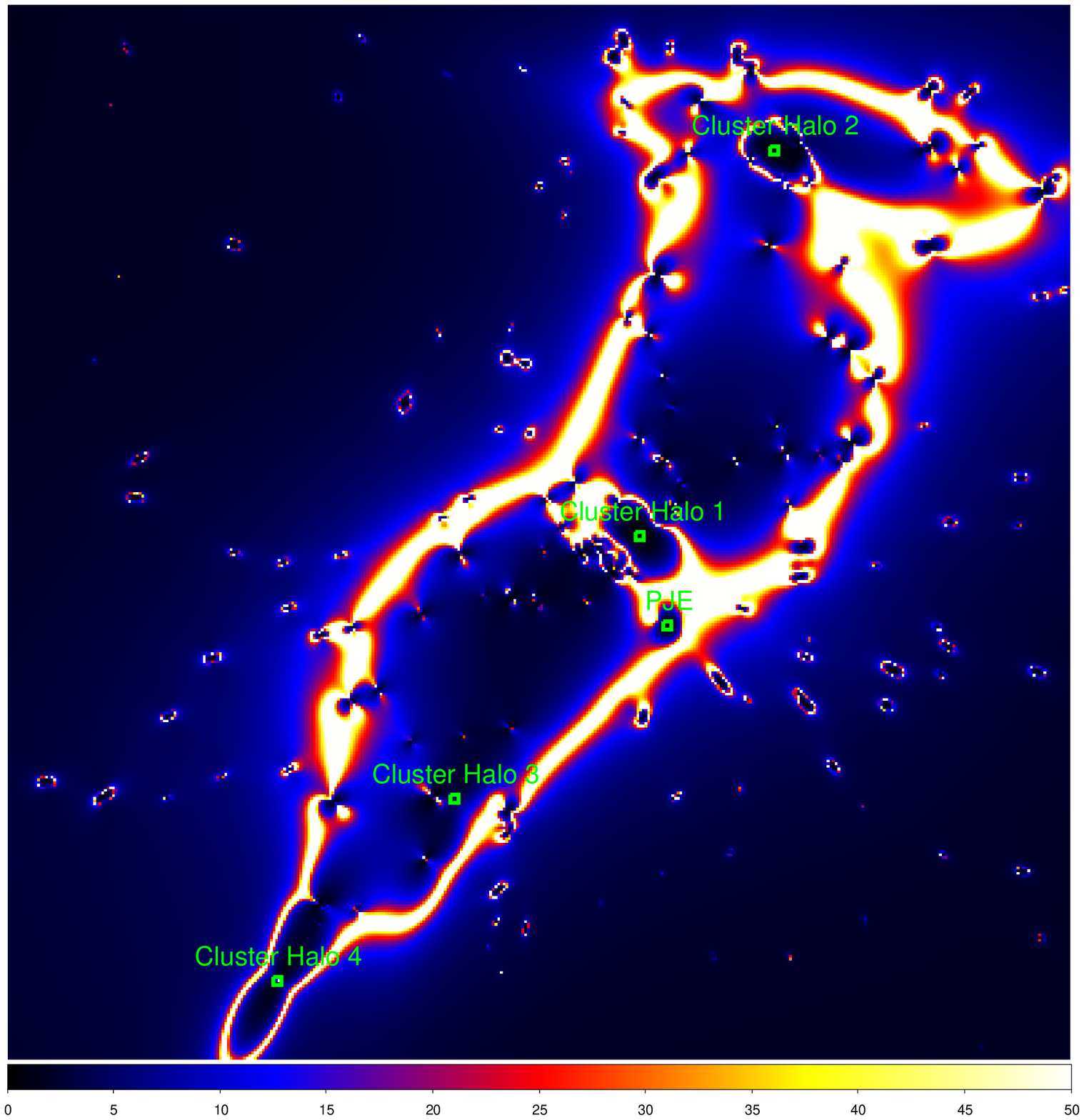}
  \caption{Positions of model components are shown on a magnification map
  for $z=9$ sources for \clone\ ({\it upper left}), \cltwo\ ({\it upper right}),
  \clthree\ ({\it lower left}), and \clfour\ ({\it lower right}).
   }
  \label{fig:magnification}
\end{figure*}

\begin{figure*}[tbp]
  \centering
      \includegraphics[width=0.45\linewidth]{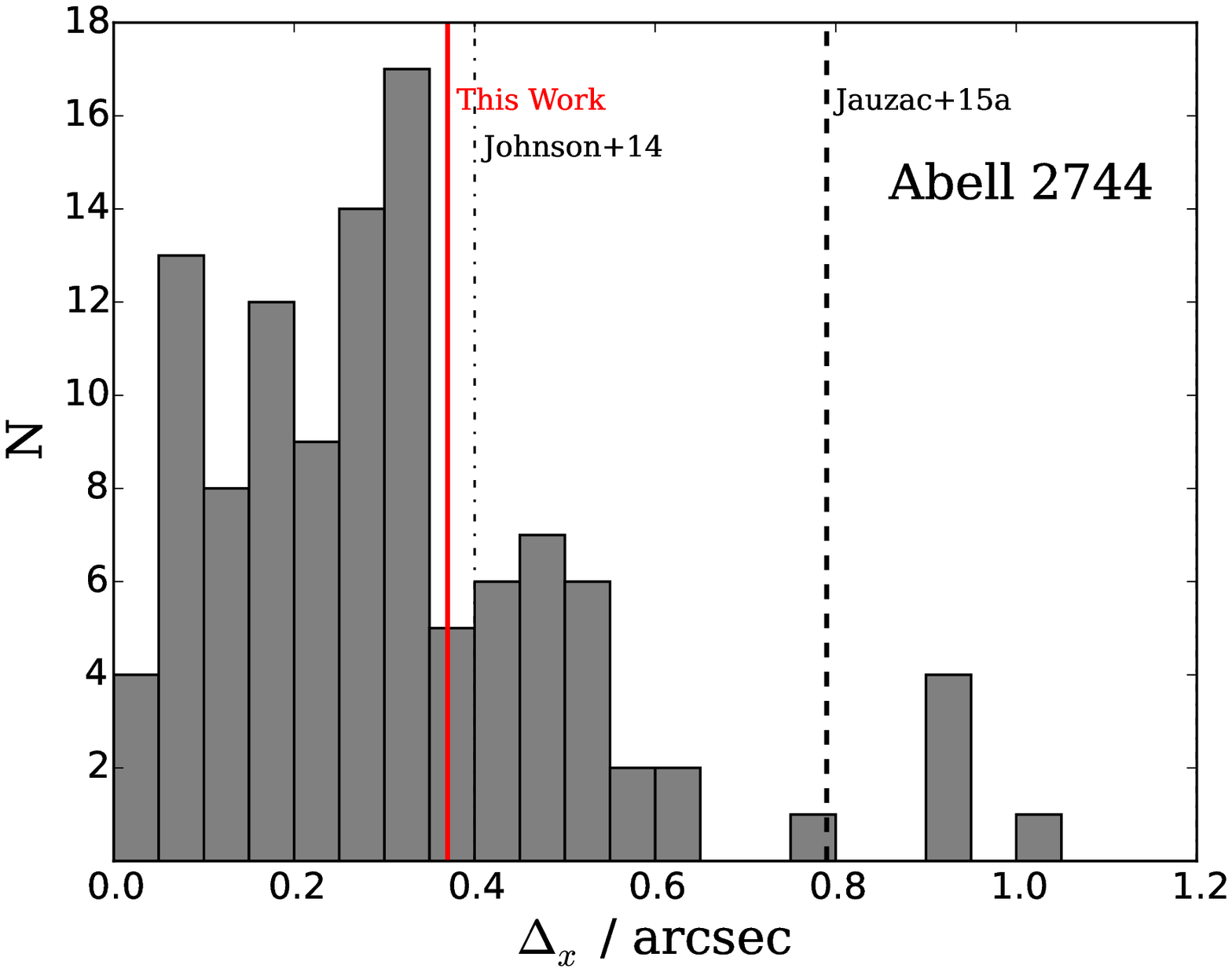}
      \includegraphics[width=0.45\linewidth]{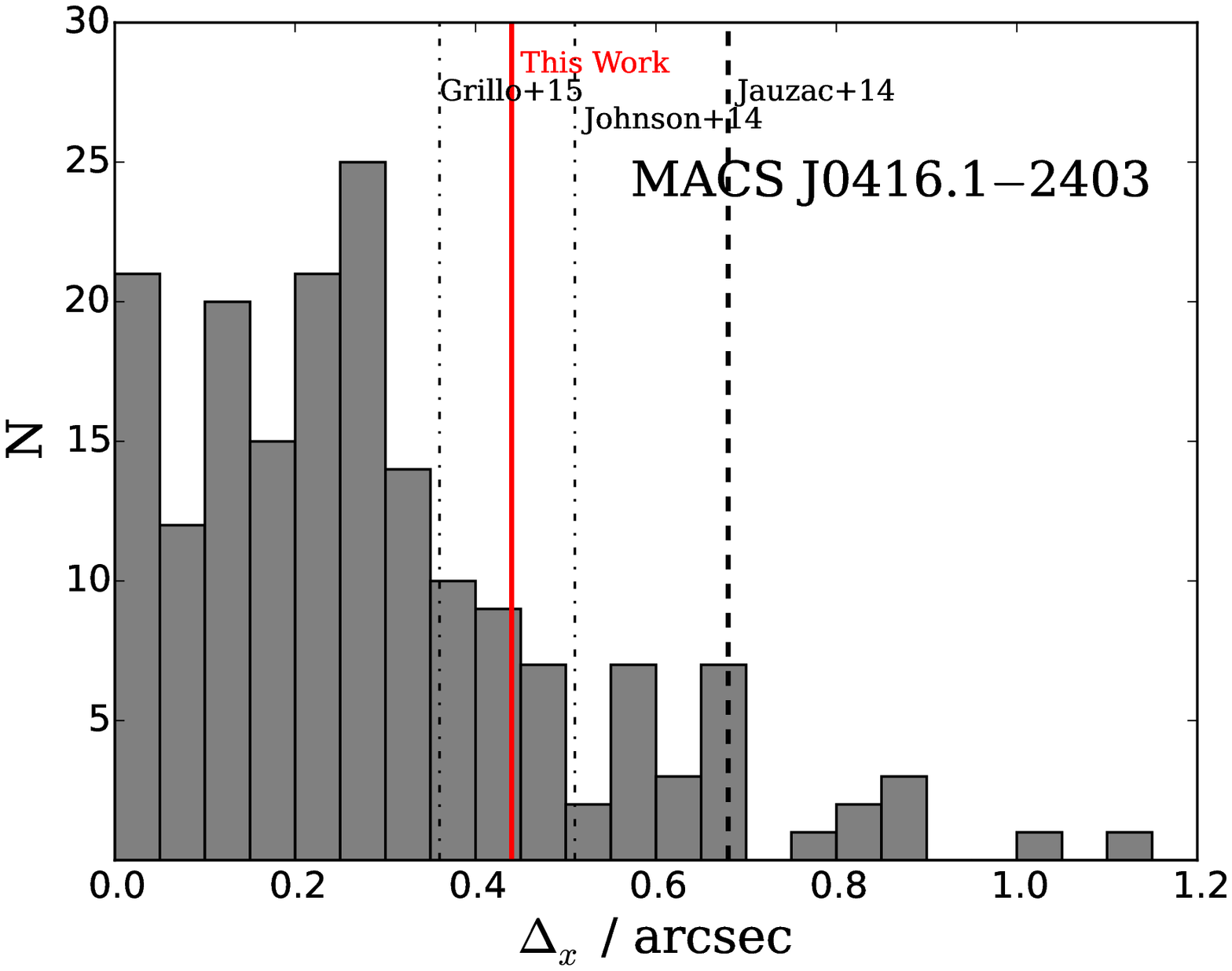}
      \includegraphics[width=0.45\linewidth]{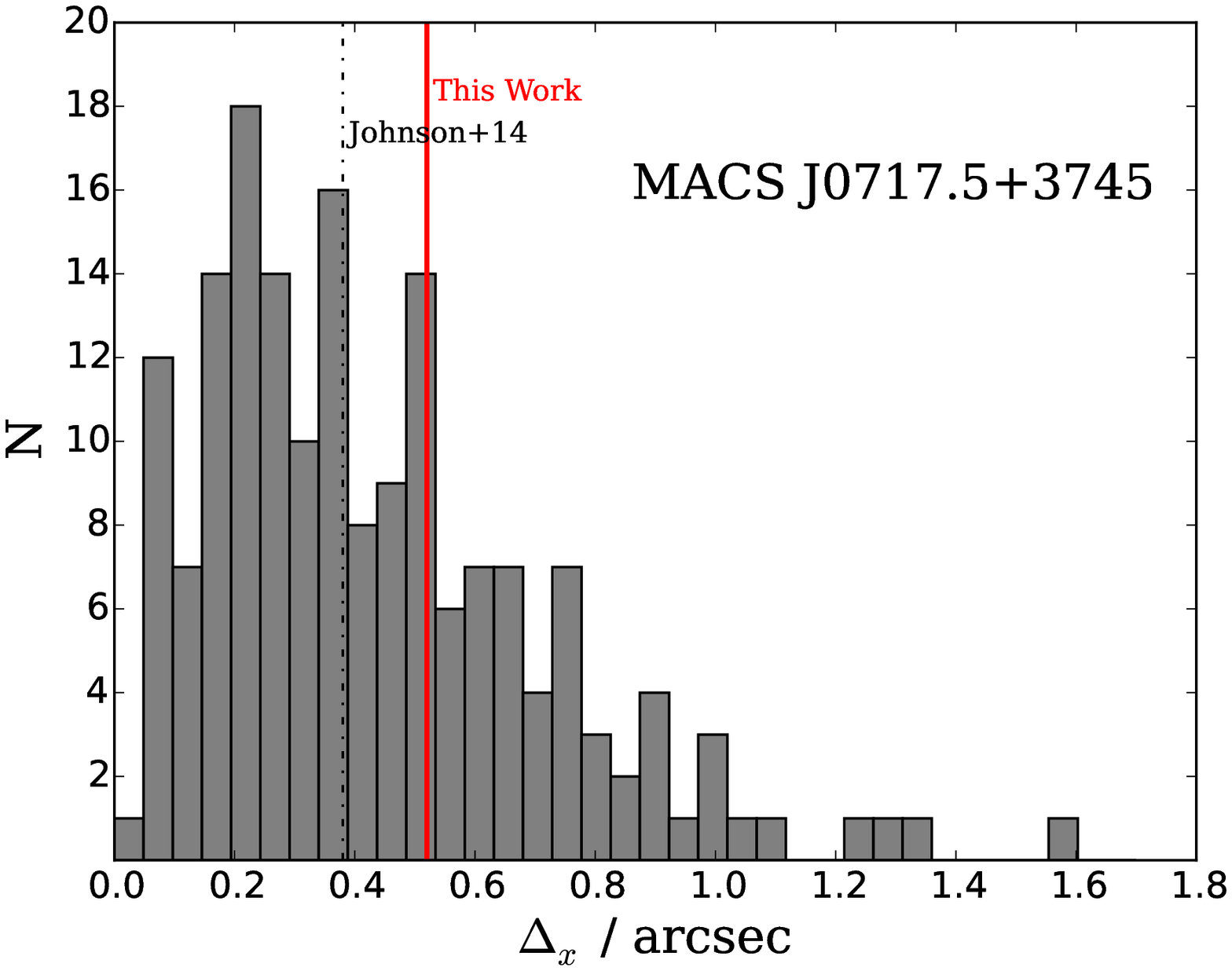}
      \includegraphics[width=0.45\linewidth]{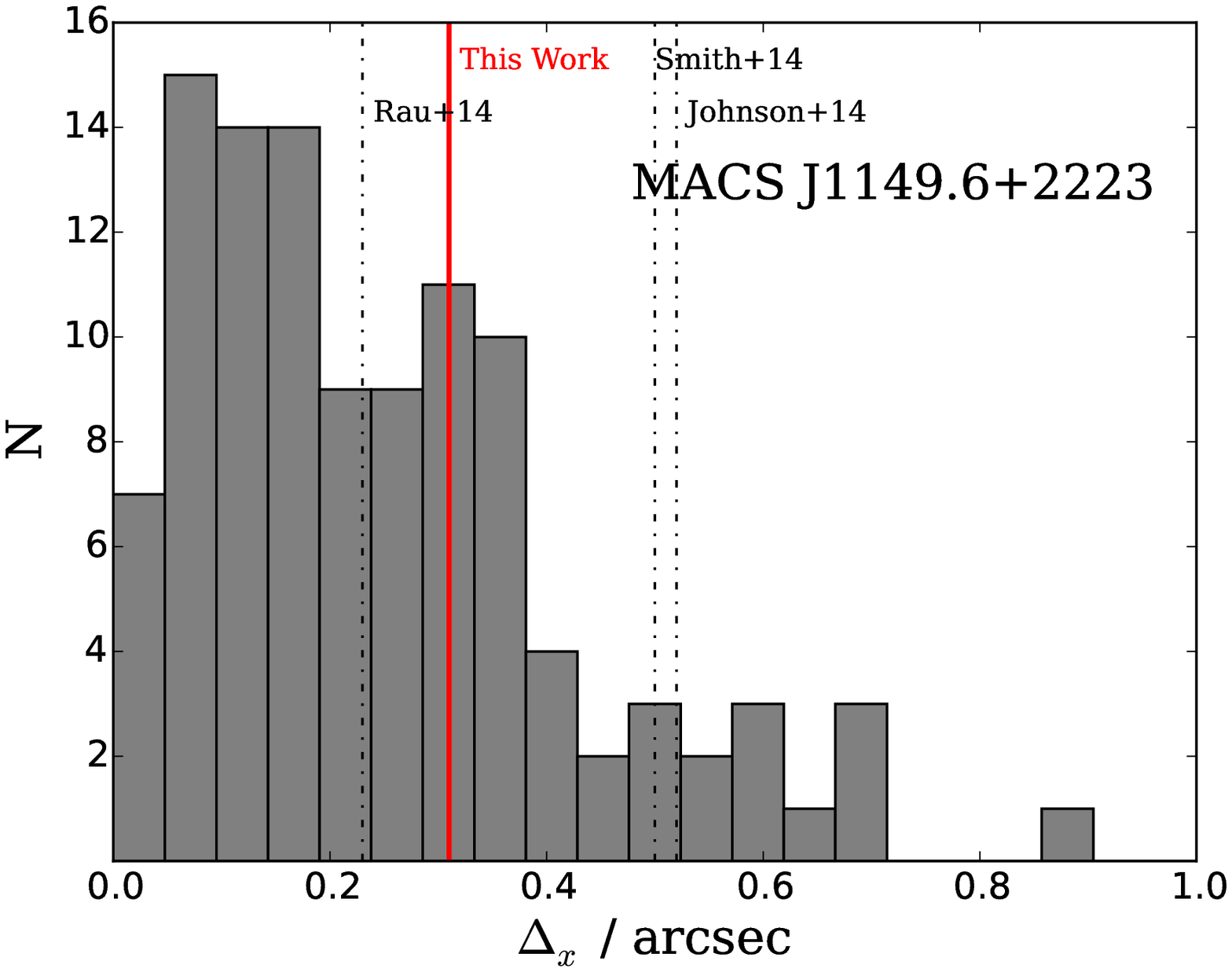}
  \caption{The distribution of the distances between observed and 
  model-predicted image positions, $\Delta_x\equiv|\mathbf{x}_{\rm
  obs}-\mathbf{x}_{\rm model}|$, for all the multiple images used for
mass modeling for \clone\ ({\it upper left}), \cltwo\ ({\it upper right}),
  \clthree\ ({\it lower left}), and \clfour\ ({\it lower right}).
See Appendix~\ref{sec:appendix_multiples} for lists of multiple
images for individual clusters. 
The red solid, black long-dashed, and black dash-dotted
 vertical lines show RMSs of $\Delta_{x}$ calculated from our models,
  previous mass models that used more than 100 multiple
  images, and previous mass models that used less than 100 multiple
  images, respectively.
The RMSs of $\Delta_x$ for all the clusters are summarized in Table~\ref{tab:modelsummaries}.}
  \label{fig:posdiff_distribution}
\end{figure*}

\subsection{The best-fitting mass models}
The numbers of input multiple images and mass modeling results of the
four HFF clusters are summarized in Table~\ref{tab:modelsummaries},
and the critical curves of the best-fitting models are shown in
Figure~\ref{fig:colorimage}. 
Figure~\ref{fig:magnification} shows magnification maps
for sources at $z=9$ and the positions of the NFW and PJE components.
We provide lists of all multiple images
used as constraints in Appendix~\ref{sec:appendix_multiples}.
Model parameters and errors from the MCMC
for individual clusters are shown in
Tables~\ref{tab:modelparams_a2744}$-$\ref{tab:modelparams_m1149}.
Parameters in square brackets are fixed during the model optimization.
Maps of magnification factor, lens potential, kappa, and shear 
  from our mass modeling will be made available on the STScI
  website\footnote{\url{https://archive.stsci.edu/prepds/frontier/lensmodels/}}.
 
Table~\ref{tab:modelsummaries} indicates that all of our best-fitting
models have reduced chi-square values, $\chi^2$/dof,
close to unity. In fact this is expected, because we have chosen 
the positional errors of multiple
images to reproduce $\chi^2/{\rm dof}\sim 1$ (see
Section~\ref{sec:input} for the specific values). In cluster strong
lensing modeling, the positional errors usually originate from the
complexity of the lens potential due to, e.g., substructures that is not
included by a simply parametrized model, rather than from
measurement uncertainties in multiple image positions. A proper
choice of positional uncertainties is important for the MCMC to
estimate model uncertainties. 

It is found that our best-fitting mass models reproduce the positions of
multiple images with RMS errors of $\sim 0\farcs4$ (see
Table~\ref{tab:modelsummaries}), which is a significant improvement over
previous strong lens modeling \citep[e.g.,][]{broadhurst05} and is
comparable or even better than other mass models constructed for
HFF. For instance, this number should be compared with RMS errors of
$0\farcs 68$ for \cltwo\ \citep{jauzac14} and $0\farcs 79$
for \clone\ \citep{jauzac15} by the CATS team, both of which used more than
100 multiple images as constraints.  \citet{grillo15} modeled
\cltwo\ with RMS errors of $0\farcs36$, but only 30 multiple images
were used as constraints. Our mass modeling satisfies both a large
number of  multiple images and a good accuracy in their reproduced
positions.  

To illustrate this point further, in Figure~\ref{fig:posdiff_distribution}
we plot the distributions of $\Delta_x\equiv|\mathbf{x}_{\rm
  obs}-\mathbf{x}_{\rm model}|$, the
distance between the observed and model-predicted image positions for
each multiple image. We find that for any cluster $\Delta_x$ is indeed 
small for most of the multiple images, with a distribution peaking 
around $0\farcs2$ and most multiple images having $\Delta_x<0\farcs6$, 
which again indicates the success of our mass modeling. 

The accuracy of our mass models may be tested further by observations of
other than image positions. For \clone, our model yields a
magnification $\mu=2.26\pm 0.12$ at the position of the lensed Type Ia
supernova HFF14Tom \citep{rodney15}. This is fully consistent with the
observed magnification $\mu=2.03\pm0.29$, although we note that this
may not be a fair comparison as we have explicitly included the observed
magnification as a constraint in mass modeling. On the other hand, the
time delays and flux ratios of the lensed supernova SN Refsdal
\citep{kelly15} in \clfour\ can provide a useful blind test of our mass
model. 
We will discuss this blind test in Section~\ref{sec:SNprediction}.

As shown in Tables~\ref{tab:modelparams_a2744}$-$\ref{tab:modelparams_m1149},
some NFW components have high ellipticities ($e > 0.7$).
There are presumably two reasons for this.
The first reason is that the intrinsic mass distribution is indeed
highly elongated, which is not surprising given the axis-ratio 
distribution of simulated dark matter halos \citep[e.g.,][]{jing02}.
In some cases, such as Cluster halos 2 and 9 in \clthree, 
such a high elongation is also 
implied by aligned positions of nearby member galaxies.
The second reason may be an insufficient number of
multiple images around the position of the NFW component.
If multiple images are unevenly distributed around an NFW
component, the model parameters can sometimes be biased toward 
the local potential, around where the multiple images are observed.
This is the case for Cluster halo 3 in the \clone\ field and Cluster 
halo 8 in the \clthree\ field.
In the case of Cluster halo 3 in the \clone\ field,
an additional NFW component is required besides the GALS component
so that the positions of the multiple images located 20 arcsec 
southwest are well reproduced.
However, this component is optimized to have a higher ellipticity
than the actual galaxy light distribution
presumably because of the small number of multiple 
images around it to constrain its parameters.

\subsection{Model comparison}
Some teams have also constructed precise mass models exploiting 
the full-depth HFF data and more than 100 multiple images.
We here compare our best-fitting mass models with those 
obtained in previous work.

{\it \clone} ---
We place three cluster-scale NFW components to model the cluster mass 
distribution.
The positions of Cluster halos 1 and 2 are consistent with 
those in \citet{jauzac15}.
\citet{wang15}, who adopt 
a free-form modeling method, also predict mass peaks 
at these positions.
In addition, we assume a third NFW component, Cluster halo 3,
as described above, where there is also a mass peak in 
\citeauthor{wang15}'s \citeyearpar{wang15} model.

{\it \cltwo} ---
We place three cluster-scale NFW components and one PJE component.
The PJE component is for better modeling of the member galaxy 
near systems 1, 2, 6, 89, and 90, as this member has a significant 
effect on these multiple image systems.
The positions of Cluster halos 1 and 2 are consistent with 
those in \citet{jauzac14} and \citet{diego15a}, but the PJE component
is included only in our model.
While \citet{jauzac14} and \citet{diego15a} assume only two halo
components, there is a ``soft component'' in the model of \citet{diego15a}
at the position of our Cluster halo 3.

{\it \clthree} ---
\citet{limousin15} use four halo-scale profiles. \citet{diego15b} 
also identify four mass peaks in their free-form model.
While we place nine cluster-scale NFW components, only four,
Cluster halos 1+3, 2, 4, and 5, have a significant mass peak.
This is consistent with their results.
\citet{limousin15} report very shallow mass profiles for this cluster, which is 
consistent with our NFW components having relatively smaller concentration parameters.
We note that the position of Cluster halo 9 is consistent with an X-ray emission peak
from {\it Chandra} \citep[see Figure 4 in][]{diego15b}.

{\it \clfour} ---
We place four cluster-scale NFW components and one PJE component.
The positions of Cluster halos 1, 2, and 3 are consistent with 
those in \citet{jauzac16}.
They do not place a component at the position of Cluster halo 4.
On the other hand, they place a halo component at the position 
of a bright member galaxy located $\sim 100$ arcsec
north from the BCG and is out of the region of the HFF WFC3/IR observation.

\subsection{Predictions for SN Refsdal}\label{sec:SNprediction}

In our mass modeling of \clfour, we only use positions of the 
multiple images S1--S4 of SN Refsdal as observational constraints. 
Importantly, when our mass modeling was completed, 
any relative time delays and magnifications had not been measured yet, 
which indicates that observations of relative time delays and 
magnifications serve as an important {\it blind} test of our mass model.
\citet{treu15b} made a detailed comparison of predictions of 
our best-fitting model (corresponding to the short name ``Ogu-a'' in 
\citealt{treu15b}) with those from other mass modeling teams. 
\citet{treu15b} also compared predictions of relative magnifications 
and time delays between images S1--S4 with preliminary measurements, 
finding a good agreement between our best-fit model predictions and observations. 
Updated measurements and comparisons are available in \citet{rodney16}.

Most mass models of \clfour\ predict two additional images of SN Refsdal 
around images 1.2 and 1.3, which we call SX and SY following \citet{oguri15}. 
SX is predicted to appear approximately one year after S1--S4, 
whereas SY is predicted to have appeared a decade ago. 
Our refined model predictions for
the time delay, position, and magnification factor of SX are
$\Delta t_{\rm SX} = 336^{+22}_{-20}$ days, $x_{\rm SX}=-4.16^{+0.08}_{-0.07}$ arcsec, 
$y_{\rm SX}=-6.50^{+0.08}_{-0.08}$ arcsec, and $\mu_{\rm SX}=4.23^{+0.32}_{-0.31}$,
where $\Delta t_{\rm SX}$ is the relative time delay from the image S1,
$x_{\rm SX}$ and $y_{\rm SX}$ are coordinates relative to the BCG.
The predicted time delay, position, and magnification factor of SY are
$\Delta t_{\rm SY} = -6229^{+209}_{-227}$ days, $x_{\rm SY}=-16.7^{+0.08}_{-0.08}$ arcsec, 
$y_{\rm SY}=12.8^{+0.12}_{-0.12}$ arcsec, and $\mu_{\rm SY}=3.52^{+0.19}_{-0.17}$.

While this paper is under review, a new SN image was discovered 
in {\it HST} images taken on 11 December \citep{kelly16}.
The observed position of the image is $x=-4.43$ arcsec and $y=-6.62$ arcsec, 
which is fully consistent with the predicted position of SX with offsets
from the predicted position only $0.27$ arcsec to the east and $0.12$ 
arcsec to the south.
Furthermore, as can be seen in Figure 2 in \citet{kelly16}, 
our time delay and magnification predictions on SX are 
fully consistent with the observed values. 
We again emphasize that these predictions are 
made before the reappearance of the new image. 
These blind test results support the validity and 
accuracy of our mass modeling method.

\section{Dropout Galaxy Sample}\label{sec:dropout}

\subsection{Lyman Break Galaxy Selection}

Galaxies at $z\sim6-7$ are selected by the Lyman break technique 
with the continuum spectral break falling in the \iFilter\ band. 
We adopt the selection criteria used in \citet{atek15}
\begin{eqnarray}
i_{814} - Y_{105} &>& 0.8,\\
Y_{105} - J_{125} &<& 0.8,\\
i_{814} - Y_{105} &>& 2(Y_{105} - J_{125}) + 0.6.
\end{eqnarray}
Objects which show $2\sigma$ level signals in both the
\bFilter\ and \vFilter\ band images or in the
\bFilter+\vFilter\ stacked image are excluded.
We require that objects need to be detected at the $5\sigma$ level 
both in the \yFilter\ and \jFilter\ bands. 
For an object not detected in the \iFilter\ band, 
we calculate the $i_{814} - Y_{105}$ color assigning
the 2$\sigma$ limiting magnitude to the \iFilter\ band magnitude.

To select galaxies at $z\sim 8$, we adopt the selection criteria 
presented in \citet{atek14}
\begin{eqnarray}
Y_{105} - J_{125} &>& 0.5,\\
J_{125} - J\!H_{140} &<& 0.5,\\
Y_{105} - J_{125} &>& 0.4 + 1.6(J_{125} - J\!H_{140}).
\end{eqnarray}
Objects which show a $2\sigma$ level signal in at least one of
the \bFilter, \vFilter, or \iFilter\ band image are excluded.
Again, objects also need to be detected at the $5\sigma$ level 
in all the \jFilter, \jhFilter, and \hFilter\ band images.

To select galaxies at $z\sim 9$, we adopt the selection criteria 
similar to those presented in \citet{ishigaki15}
\begin{eqnarray}
(Y_{105} + J_{125})/2 - J\!H_{140} &>& 0.75,\\
(Y_{105} + J_{125})/2 - J\!H_{140} \nonumber\\
 > 0.75 &+& 0.8 (J\!H_{140} - H_{160}),\\
J_{125} - H_{160} &<& 1.15,\\
J\!H_{140} - H_{160} &<& 0.6.
\end{eqnarray}
Objects which show a $2\sigma$ level signal 
in at least one of the \bFilter, \vFilter, or \iFilter\ band image are
excluded. We require that objects need to be detected at the $3\sigma$
level in both the \jhFilter\ and \hFilter\ band images and at the
$3.5\sigma$ level in at least one of these two bands.
If an object is fainter than the 0.9$\sigma$ level magnitude
in the \yFilter\ or \jFilter\ band,
we assign the 0.9$\sigma$ level magnitude 
to the photometry of that band.

In addition, we adopt a pseudo-$\chi^{2}$ constraint to reduce the
contamination rate. This constraint is defined as $\chi_{\rm opt}^{2}
< 2.8$, where $\chi_{\rm opt}^{2} = \sum_{i} {\rm SGN} (f_{i}) (f_{i}
/ \sigma_{i})^{2}$.
Here, $f_{i}$ is the flux density in the $i$-th band and ${\rm SGN}(x)$ is the sign function defined by
  ${\rm SGN}(x) = 1$ if $x>0$ and ${\rm SGN}(x) = -1$ if $x<0$.
The summation runs over all the optical bands.
Finally, we visually inspect all the dropout galaxy candidates and
remove seven obvious spurious sources.

\begin{figure}[tbp]
  \centering
      \includegraphics[width=1.0\linewidth, trim=20 0 30 30, clip]{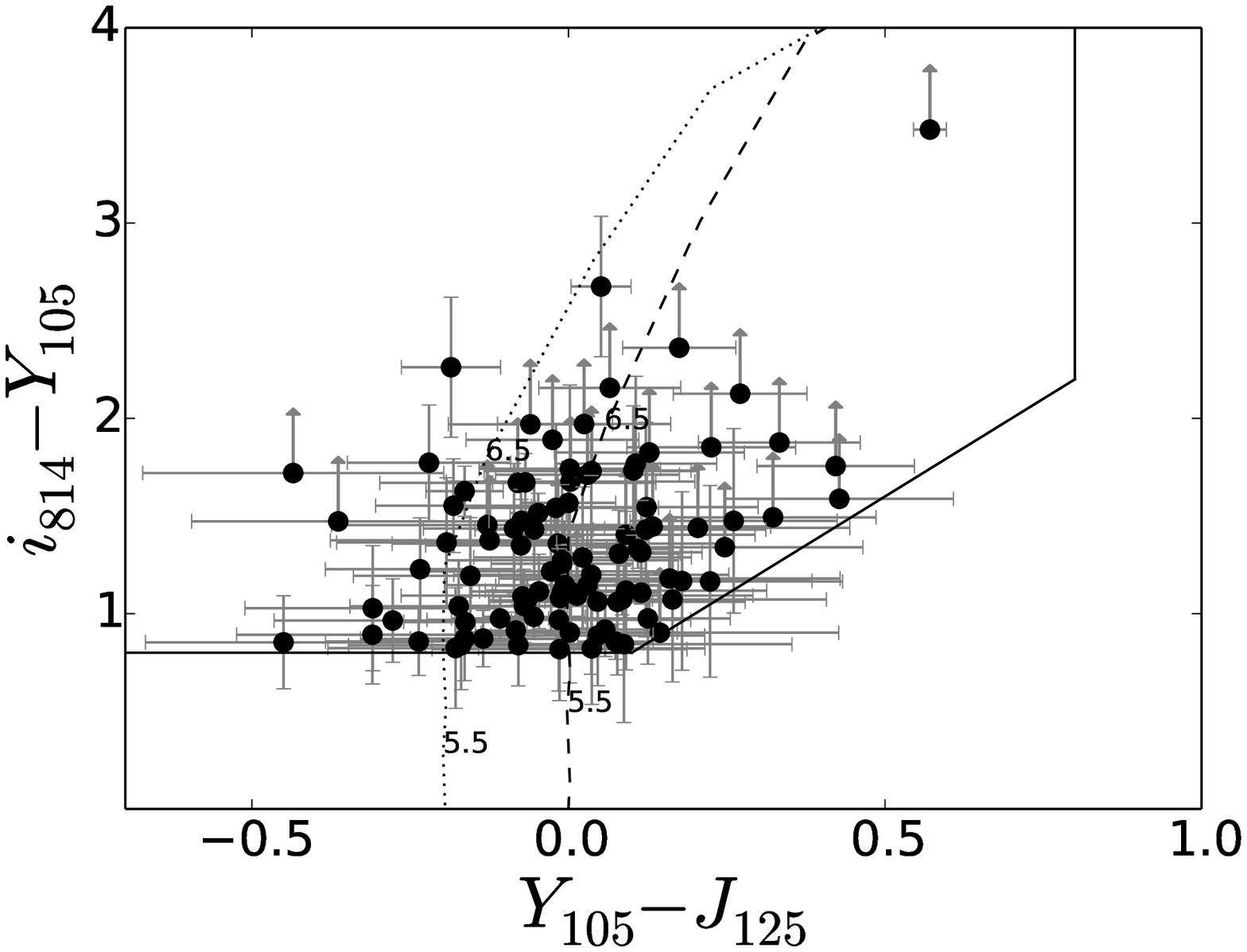}
      \includegraphics[width=1.0\linewidth, trim=20 0 30 30, clip]{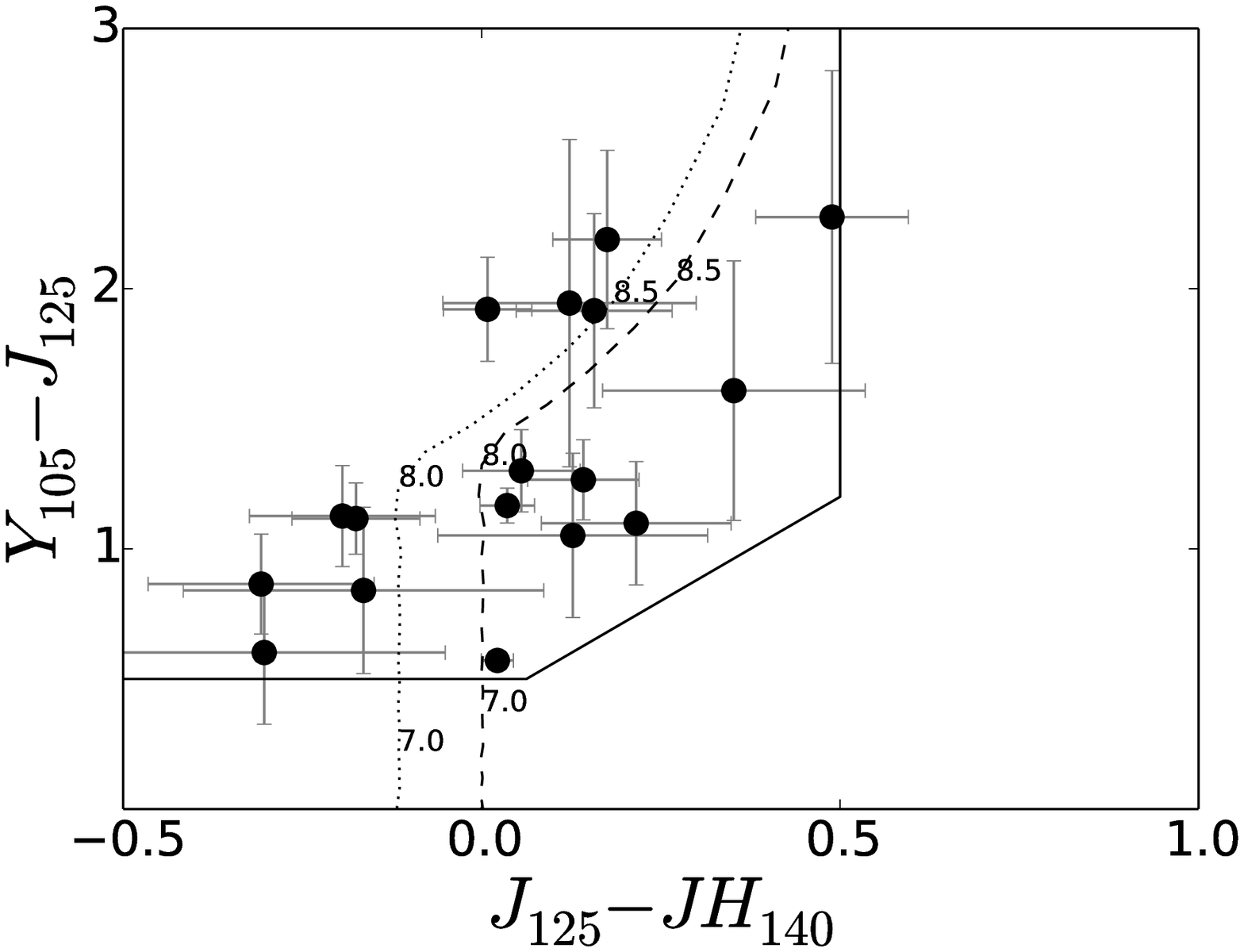}
      \includegraphics[width=1.0\linewidth, trim=20 0 30 30, clip]{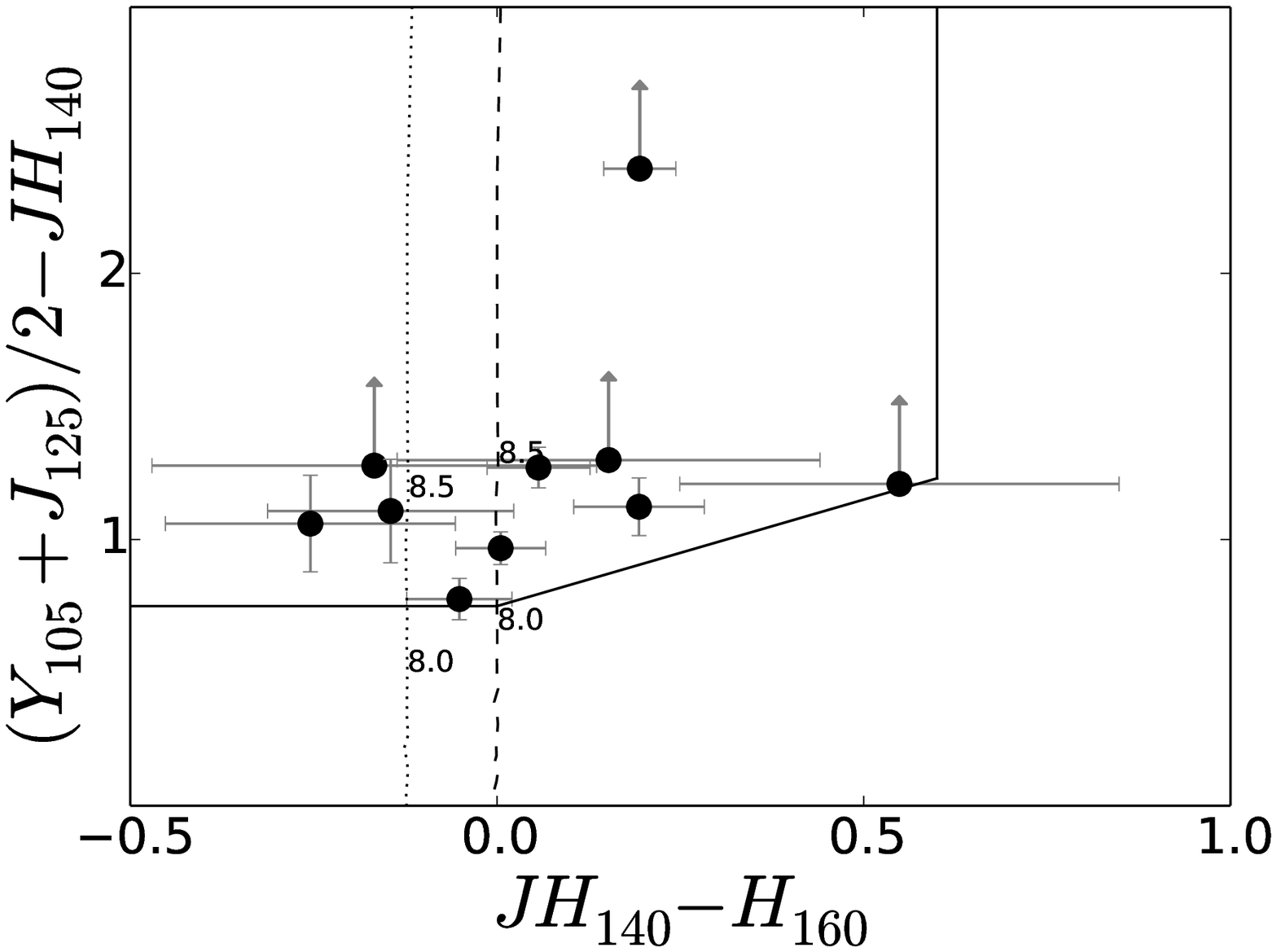}
  \caption{Two-color diagrams for \idrop\ ({\it top}), 
  \ydrop\ ({\it middle}), and \yjdrop\ ({\it bottom}) galaxy candidates,
  with our color selections indicated with solid lines.
  Filled circles represent dropout galaxy candidates.
  Tracks of expected galaxy colors computed assuming UV-slopes of
  $\beta=-2$ and $-3$ are shown with dashed and dotted lines, respectively.
  Small numbers represent assumed redshifts.
The colors of all the dropout galaxy candidates are summarized in
Tables~\ref{tab:candidates7}--\ref{tab:candidates9}.}
  \label{fig:color-color}
\end{figure}

\subsection{Dropout galaxy sample}

We list the \idrop\ ($z\sim 6-7$), \ydrop\ ($z\sim 8$), and
\yjdrop\ ($z\sim 9$) galaxies from the four HFF cluster fields in
Tables~\ref{tab:candidates7}$-$\ref{tab:candidates9} respectively
in Appendix~\ref{sec:appendix_dropouts}. 
We show the distribution of these dropout galaxies in
  color-color spaces in Figure~\ref{fig:color-color}.
For each galaxy, the first part of ID represents
  the field in which it is found; 1C, 2C, 3C,
  and 4C indicate \clone\ cluster,  \cltwo\ cluster,
  \clthree\ cluster, and \clfour\ cluster fields, respectively.
  The second part of ID represents its coordinates.
\footnote{For example, HFF1C-2251-4556 is found in \clone\ cluster field
and its coordinates are R.A.$=$00:14:22.51, Decl.$=-$30:24:55.6.}
In the Tables we also provide
magnification factors at the positions of galaxies predicted by
our mass models presented in Section~\ref{sec:result}.

In summary, we select 100 \idrop, 17 \ydrop, and 10
\yjdrop\ galaxies. Note that there are some overlaps in the dropout
samples.
We find that one object is identified by the $Y$- and \idrop\ selections, 
and that six objects meet the criteria of the $Y\!J$- and \ydrop\ selections.
Most of the dropout galaxies have a modest magnification factor, 
$\mu \la 5$, while some are highly magnified. 
For instance, based on the magnification maps of our
best-fitting models, 14 and four galaxies at 
$z\sim6-7$ and $9$, respectively, have a magnification factor
larger than 10.  Among them four at $z\sim 6-7$ and one at
$z\sim 9$  a have magnification factor larger than 50, albeit with large 
uncertainties.

\begin{figure}[tbp]
  \centering
      \includegraphics[width=\linewidth]{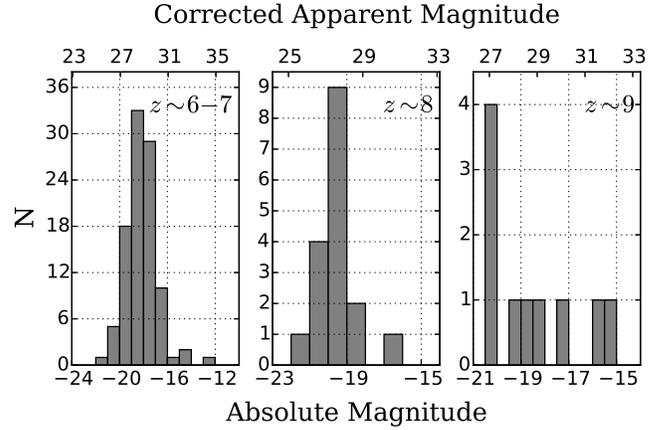}
  \caption{Histograms of dropout galaxies at $z\sim 6-7$ ({\it left
      panel}), $z\sim 8$ ({\it middle panel}), and $z\sim 9$ ({\it
      right panel}) in the four HFF cluster fields as a function of
    the intrinsic (unlensed) absolute magnitude. Magnification factors of
    individual dropout galaxies are corrected based on our
    best-fitting mass models. 
    Most of the intrinsically faint galaxies are highly magnified
    and their estimated magnitudes are affected by the
    uncertainty in the magnification factor.
    Nevertheless, magnitude errors propagated from 
    errors in the magnification factor in the $z\sim6-7$, 8, and 9 samples
    are no larger than only 0.87, 0.11, and 1.19 mag, respectively.
    Details of these dropout galaxies
  are given in Tables~\ref{tab:candidates7}$-$\ref{tab:candidates9}.
  Note that all of the dropout galaxy candidates are plotted here.}
  \label{fig:magnitude_distribution}
\end{figure}

Some of these high-magnification galaxies may be intrinsically faint. 
To examine this possibility, we plot the
histograms of all dropout galaxies as a function of
intrinsic magnitude corrected for magnification factor in
Figure~\ref{fig:magnitude_distribution}. We find that they
 typically have absolute magnitudes of $M_{\rm UV} \sim -18$, or 
 intrinsic magnitude of $\sim 29$ mag, but some are as faint as 
 $M_{\rm UV} \sim -14$, or intrinsic magnitude of $\sim 33$ mag.

\subsection{Multiple image candidates}

Our analysis suggests that some dropout galaxies are multiply
imaged. Among them, reliable ones have been included in our mass
modeling; systems 28, 46, and 54 in \clone\ field; 
systems 6, 90, 91, and 92 in \cltwo\ field; 
systems 19 and 66 in \clthree\ field;
and systems 33, 38, and 39 in \clfour\ field (see Section~\ref{sec:result}). 
Here we discuss several
interesting reliable multiple images and multiple image candidates
at $z\sim9$.

{\it HFF2C-i2, -i3, -i7, and -i16} --- These are newly identified multiple images
in \cltwo\ field. HFF2C-i2 and -i16 compose system 91, and HFF2C-i3 
and -i7 compose system 92. They are placed in the most 
northeast part of this cluster and improve mass modeling in this
region.

\begin{figure}[tbp]
  \centering
      \includegraphics[width=0.49\linewidth, trim=100 100 100 100, clip]{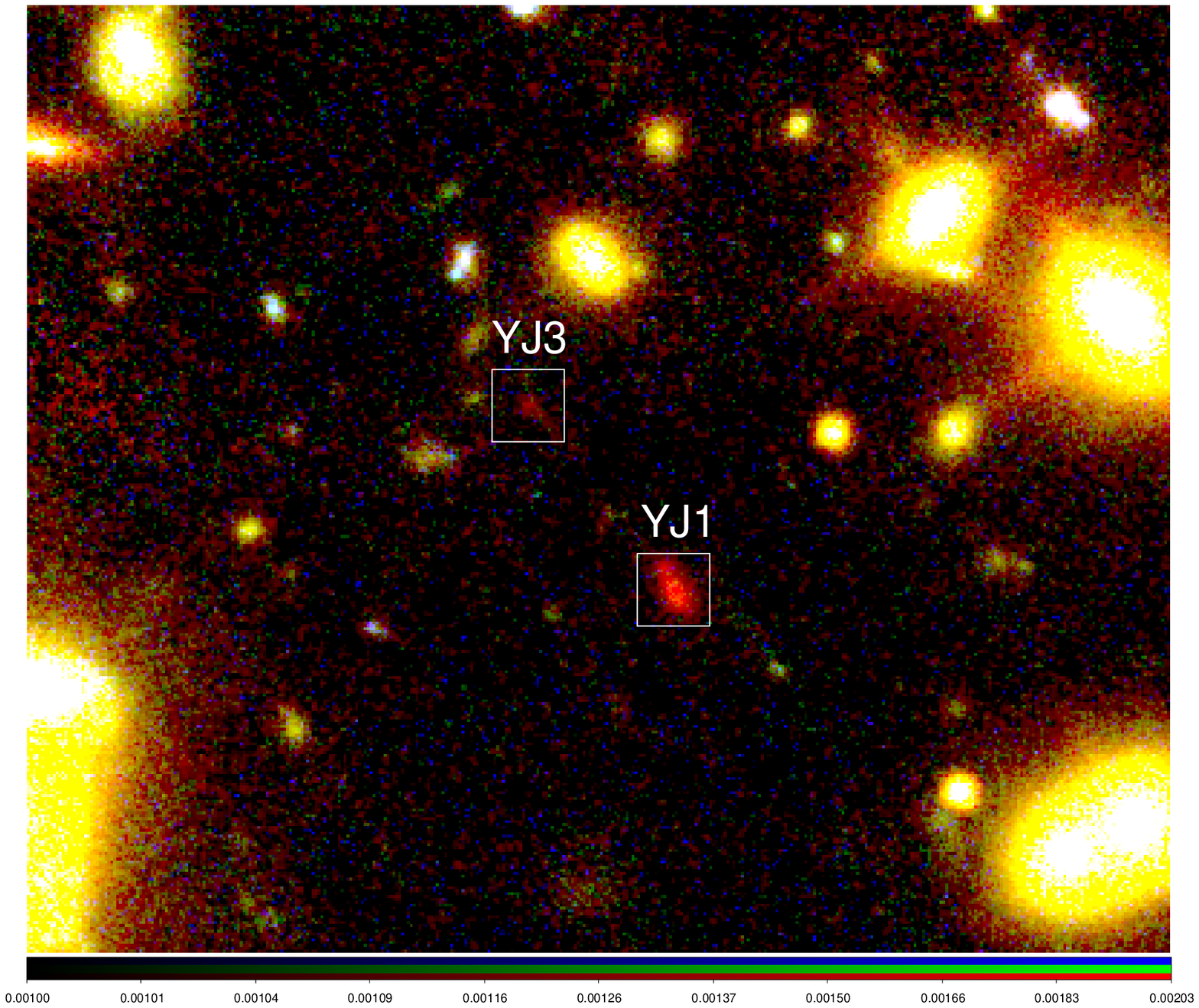}
      \includegraphics[width=0.49\linewidth, trim=100 100 100 100, clip]{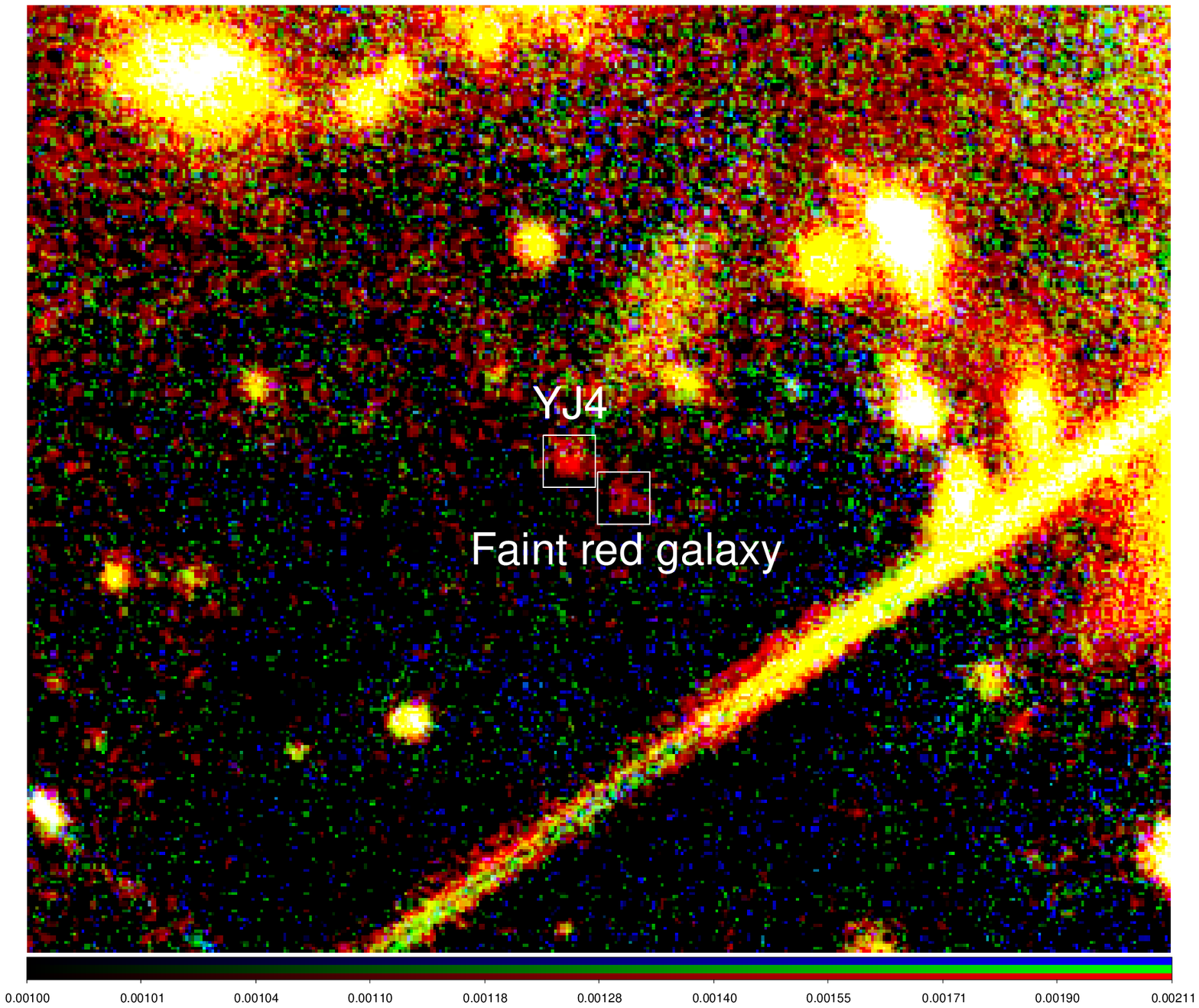}
  \caption{Color-composite images of the multiple image candidates, 
  HFF4C-YJ1 and HFF4C-YJ3 ({\it left panel}), and of HFF4C-YJ4 and its companion ({\it right panel}).
  {\it Left}: HFF4C-YJ1 and HFF4C-YJ3 may be distorted in the
  direction of the shear at that position. 
  {\it Right}: A faint red galaxy is located very close to HFF4C-YJ4, and
  its position is consistent with being a counter image of HFF4C-YJ4.
  }
  \label{fig:multiple_candidates}
\end{figure}

{\it HFF4C-YJ1 and HFF4C-YJ3} --- HFF4C-YJ1 is a bright $z\sim 9$
galaxy candidate in \clfour\ discovered by \citet{zheng12}. We find
a faint $z\sim 9$ galaxy candidate, HFF4C-YJ3, close to 
this galaxy (see Figures~\ref{fig:colorimage} and \ref{fig:multiple_candidates}). 
Our best-fitting mass model has a critical curve
that is placed near these galaxies. Therefore, it is possible that
these two galaxies are in fact multiple images of a $z\sim 9$
galaxy. The reliability of this multiple image system is not very
high, because there are not many known multiple images around this system, 
and therefore our mass model in this region includes relatively large
uncertainties. 
The \jhFilter $-$ \hFilter\ colors of YJ1 and YJ3 are $0.24\pm0.04$
and $0.16\pm0.23$, respectively, and are consistent with being
multiple images.

{\it HFF4C-YJ4} --- This is a $z\sim 9$ galaxy in \clfour\ near the
critical curve. We find another faint red galaxy nearby this galaxy
(see Figure~\ref{fig:multiple_candidates}). 
The color of this faint red galaxy is consistent with being at
$z\sim 9$, but it is below the detection limit used for the dropout selection. 
The relative positions of
these two galaxies are fully consistent with being multiple images
of a single $z\sim 9$ galaxy. Given its high reliability, we include
the positions of these galaxies as constraints in our mass modeling
as system 38.
The \jhFilter $-$ \hFilter\ colors of YJ4 and the faint red galaxy are $-0.15\pm0.24$
and $0.19\pm0.24$, respectively.
This is consistent with being multiple images.

Even if these galaxy pairs are not real multiple images of 
single galaxies, the close separations are interesting in term of
galaxy formation and evolution.

\subsection{Future Analyses}
We have presented a sample of high-redshift galaxies selected in 
the cluster fields where lensing effects are significant.
If those from the accompanied parallel fields are added,
we will have a four times larger sample of $z\gtrsim6$ galaxies
than that used in our previous studies \citep{ishigaki15, kawamata15a}.
In the forthcoming papers, we plan to use this large sample
to investigate various properties of high-redshift galaxies
including luminosity functions, sizes and morphologies, and stellar populations,
and their implications for cosmic reionization.

The absolute magnitudes of the new high-redshift galaxy sample 
constructed in this paper
extend down to $M_{\rm UV} \simeq -12.1$, $-16.8$, and 
$-15.0$ at $z\sim6-7$, $8$, and $9$, 
respectively, enabling us to study extremely faint galaxies in the 
reionization era.
These limiting magnitudes at $z\sim 6-7$ and $9$ are significantly fainter than 
those in previous studies based on only one or two clusters
(e.g., $M_{\rm UV} \simeq -15.25$ at $z\sim 6-7$ in \citealt{atek16} 
and $M_{\rm UV} \simeq -18.1$ at $z\sim 9$ in \citealt{mcleod15}).

\section{Conclusion}\label{sec:conclusion}

We have conducted precise mass modeling of four HFF clusters,
exploiting the full depth HFF data and the latest spectroscopic
follow-up results on multiple images. We have used the positions of
111, 182, 173, and 108 multiple images to constrain the matter
distributions of \clone, \cltwo, \clthree, and \clfour, respectively. 
Among them, 145 multiple images are new systems identified
in this paper.  We assume simply parametrized mass models and optimize
model parameters with the public software {\sc glafic}
\citep{oguri10}. We have found that our best-fitting mass models reproduce the
observed positions of multiple images quite well, with image plane RMS
of $\sim 0\farcs 4$ (see Table~\ref{tab:modelsummaries}). For Abell
2744, our best-fitting mass model recovers the observed magnification at
the position of the Type Ia supernova HFF14Tom \citep{rodney15},
although we note that we have explicitly included this magnification as a
constraint in mass modeling. We have found that the predicted time delays
and flux ratios of the quadruple images of SN Refsdal \citep{kelly15} in
\clfour\ are consistent with observations \citep{treu15b}. 

We have then constructed $z\sim6-9$ dropout galaxy catalogs in these four
cluster fields from the full depth HFF images. For each dropout galaxy we
have estimated the magnification factor from our mass modeling results.
The catalogs consist of 100 galaxies at $z\sim 6-7$, 17 galaxies at
$z\sim 8$, and 10 galaxies at $z\sim 9$, although some of them
are detected in multiple dropout selections. While most of these
galaxies have modest magnifications, $\mu \la 5$, there are several
dropout galaxies with a magnification larger than 10. Specifically, 14
at $z\sim 6-7$ and four at $z\sim 9$ have a predicted 
magnification factor larger than 10. 
The intrinsic (unlensed) magnitudes of these high-magnification galaxies are
estimated to be $\sim 32-33$ mag, which indicates that the HFF program
indeed detects the faintest galaxies known to date.  

%%%%%%%%%%%%%%%%%%%%%%%%%%%%%%%%%%%%%%%%%%%%%%%%%%%%%%%%%%%%
%%%%%%%%%%%%%%%%%%%%%%%%%%%%%%%%%%%%%%%%%%%%%%%%%%%%%%%%%%%%
\section*{Acknowledgments}
%%%%%%%%%%%%%%%%%%%%%%%%%%%%%%%%%%%%%%%%%%%%%%%%%%%%%%%%%%%%%
%%%%%%%%%%%%%%%%%%%%%%%%%%%%%%%%%%%%%%%%%%%%%%%%%%%%%%%%%%%%%
We thank the Frontier Fields mass modeling initiative led by Dan Coe and 
participants of this initiative for sharing invaluable data before the publications. 
We are grateful to the authors of \citet{treu15b} for
sharing the follow-up data of MACS J1149.6+2223 and for
helpful discussions.
This work was supported in part by World Premier International
Research Center Initiative (WPI Initiative), MEXT, Japan, and
JSPS KAKENHI Grant Number 26800093, 23244025, 15H05892, and 15H02064. 

%Facilities: \facility{HST(ACS/WFC3IR)}.

\bibliographystyle{apj}
\bibliography{bibtex.bib}

%%%%%%%%%%%%%%%%%%%%%%%%%%%%%%%%%%%%%%%%%%%%
% Multiple images
%%%%%%%%%%%%%%%%%%%%%%%%%%%%%%%%%%%%%%%%%%%%

\begin{appendix}

\section{Lists of multiple images used for mass modeling}\label{sec:appendix_multiples}

The list of multiple images we use for mass modeling is given in
Tables~\ref{tab:a2744multiple}$-$\ref{tab:m1149multiple}.

%\clearpage
\LongTables
\begin{deluxetable*}{rcccccl}
\tabletypesize{\scriptsize}
\tablecaption{\clone\ Multiple Image Systems\label{tab:a2744multiple}}
\tablewidth{0pt}
\tablehead{
      \colhead{ID} & \colhead{R.A.} & \colhead{Decl.} & \colhead{$z_{\rm spec}$} & \colhead{$z_{\rm model}$} & \colhead{Photo-z prior} & \colhead{Reference\tablenotemark{a}} 
}
\startdata
 1.1 & $3.597542$ & $-30.403917$ & 1.500 & \nodata & \nodata & JM, XW\\
 1.2 & $3.595958$ & $-30.406822$ &  & \nodata & \nodata & JM\\
 1.3 & $3.586208$ & $-30.409986$ &  & \nodata & \nodata & JM\\
 2.1 & $3.583250$ & $-30.403350$ & \nodata & $1.89^{+0.07}_{-0.07}$ & \nodata & JM\\
 2.2 & $3.597292$ & $-30.396725$ & \nodata &  & \nodata & JM\\
 2.3 & $3.585417$ & $-30.399897$ & \nodata &  & \nodata & JM\\
 2.4 & $3.586417$ & $-30.402128$ & \nodata &  & \nodata & JM\\
 3.1 & $3.589480$ & $-30.393867$ & 3.980 & \nodata & \nodata & JM, TJ\\
\enddata
\tablenotetext{a}{JM~=~\citet{merten11}, HA14~=~\citet{atek14}, JR~=~\citet{richard14}, AZ~=~\citet{zitrin14}, DL~=~\citet{lam14}, TJ~=~\citet{johnson14}, MI~=~\citet{ishigaki15}, MJ~=~\citet{jauzac15}, XW~=~\citet{wang15}.}
\end{deluxetable*}

%\clearpage
\LongTables
\begin{deluxetable*}{rcccccl}
\tabletypesize{\scriptsize}
\tablecaption{\cltwo\ Multiple Image Systems\label{tab:m0416multiple}}
\tablewidth{0pt}
\tablehead{
      \colhead{ID} & \colhead{R.A.} & \colhead{Decl.} & \colhead{$z_{\rm spec}$} & \colhead{$z_{\rm model}$} & \colhead{Photo-z prior} & \colhead{Reference\tablenotemark{a}} 
}
\startdata
 1.1 & $64.040797$ & $-24.061616$ & 1.896 & \nodata & \nodata & AZ, LC\\
 1.2 & $64.043472$ & $-24.063514$ &  & \nodata & \nodata & AZ\\
 1.3 & $64.047367$ & $-24.068673$ &  & \nodata & \nodata & AZ\\
 2.1 & $64.041168$ & $-24.061851$ & 1.893 & \nodata & \nodata & AZ, CG\\
 2.2 & $64.043021$ & $-24.063010$ &  & \nodata & \nodata & AZ\\
 2.3 & $64.047484$ & $-24.068851$ &  & \nodata & \nodata & AZ\\
 3.1 & $64.030772$ & $-24.067118$ & 1.989 & \nodata & \nodata & AZ, CG\\
 3.2 & $64.035254$ & $-24.070993$ &  & \nodata & \nodata & AZ\\
\enddata
\tablenotetext{a}{LC~=~\citet{christensen12}, AZ~=~\citet{zitrin13}, MJ~=~\citet{jauzac14}, CG~=~\citet{grillo15}, JD~=~\citet{diego15a}, AH~=~Hoag~et~al.~(in prep.), SR~=~Rodney~et~al.~(in prep.).}
\end{deluxetable*}

%\clearpage
\LongTables
\begin{deluxetable*}{rcccccl}
\tabletypesize{\scriptsize}
\tablecaption{\clthree\ Multiple Image Systems\label{tab:m0717multiple}}
\tablewidth{0pt}
\tablehead{
      \colhead{ID} & \colhead{R.A.} & \colhead{Decl.} & \colhead{$z_{\rm spec}$} & \colhead{$z_{\rm model}$} & \colhead{Photo-z prior} & \colhead{Reference\tablenotemark{a}} 
}
\startdata
  1.1 & $109.395338$ & $+37.741175$ & 2.963 & \nodata & \nodata & AZ, ML\\
 1.2 & $109.393826$ & $+37.740092$ &  & \nodata & \nodata & AZ\\
 1.3 & $109.390988$ & $+37.738286$ &  & \nodata & \nodata & AZ\\
 1.4 & $109.384352$ & $+37.736947$ &  & \nodata & \nodata & AZ\\
 1.5 & $109.405784$ & $+37.761374$ &  & \nodata & \nodata & EV\\
 2.1 & $109.392813$ & $+37.741003$ & \nodata & $2.82^{+0.18}_{-0.11}$ & \nodata & AZ\\
 2.2 & $109.390434$ & $+37.739247$ & \nodata &  & \nodata & AZ\\
 3.1 & $109.398546$ & $+37.741503$ & 1.855 & \nodata & \nodata & AZ, ML\\
\enddata
\tablenotetext{a}{AZ~=~\citet{zitrin09b}, ML~=~\citet{limousin12}, KS~=~\citet{schmidt14}, EV~=~\citet{vanzella14}, JR~=~\citet{richard14}, JD~=~\citet{diego15b}, TT~=~\citet{treu15a}}
\end{deluxetable*}

%\clearpage
\LongTables
\begin{deluxetable*}{rcccccl}
\tabletypesize{\scriptsize}
\tablecaption{\clfour\ Multiple Image Systems\label{tab:m1149multiple}}
\tablewidth{0pt}
\tablehead{
      \colhead{ID} & \colhead{R.A.} & \colhead{Decl.} & \colhead{$z_{\rm spec}$} & \colhead{$z_{\rm model}$} & \colhead{Photo-z prior} & \colhead{Reference\tablenotemark{a}} 
}
\startdata
1.1 & $177.397003$ & $+22.396003$ & 1.488 & \nodata & \nodata & AZ, GS\\
 1.2 & $177.399420$ & $+22.397442$ &  & \nodata & \nodata & AZ\\
 1.3 & $177.403420$ & $+22.402442$ &  & \nodata & \nodata & AZ\\
 2.1 & $177.402419$ & $+22.389753$ & 1.894 & \nodata & \nodata & AZ, GS\\
 2.2 & $177.406044$ & $+22.392480$ &  & \nodata & \nodata & AZ\\
 2.3 & $177.406586$ & $+22.392889$ &  & \nodata & \nodata & AZ\\
 3.1 & $177.390753$ & $+22.399850$ & 3.129 & \nodata & \nodata & AZ, MJ, CG\\
 3.2 & $177.392711$ & $+22.403083$ &  & \nodata & \nodata & AZ\\
\enddata
\tablenotetext{a}{AZ~=~\citet{zitrin09a}, GS~=~\citet{smith09}, WZ~=~\citet{zheng12}, SR~=~\citet{rau14}, JR~=~\citet{richard14}, PK~=~\citet{kelly15}, MJ~=~\citet{jauzac16}, TT~=~\citet{treu15b}, CG~=~\citet{grillo16}, GB~=~Brammer~et~al.~(in prep.).}
\tablenotetext{*}{We note that in the identification of the multiple images, we do not refer to \citet{jauzac16}, which was posted to arXiv very recently. All of the multiple images labeled \citet{jauzac16} in this table are also presented in \citet{treu15b}.}
\end{deluxetable*}

%%%%%%%%%%%%%%%%%%%%%%%%%%%%%%%%%%%%%%%%%%%%
% Dropout galaxies
%%%%%%%%%%%%%%%%%%%%%%%%%%%%%%%%%%%%%%%%%%%%

\section{Lists of dropout galaxy candidates}\label{sec:appendix_dropouts}

The list of dropout galaxy candidates is given in
Tables~\ref{tab:candidates7}$-$\ref{tab:candidates9}.

\LongTables
\begin{deluxetable*}{llcclrcrrcl} 
\tabletypesize{\scriptsize}
\tablecaption{Dropout galaxy candidates at $z\sim 6-7$\label{tab:candidates7} 
in the four HFF cluster fields.}
\tablewidth{0pt}
\tablehead{
  \colhead{ID\tablenotemark{a}} &  \colhead{ID\tablenotemark{a,b}} & \colhead{R.A.} & \colhead{Decl.} & \colhead{$i_{814}-Y_{105}$} & \colhead{$Y_{105}-J_{125}$} & \colhead{$J_{125}$\tablenotemark{b}} & \colhead{$\mu_{\rm best}$} & \colhead{$\mu$\tablenotemark{c}} & \colhead{$z_{\rm photo}$} & \colhead{Reference\tablenotemark{d}}
}
\startdata
\sidehead{HFF1C}
2251-4556 & i1 & $ 3.593804 $ & $ -30.415447 $ & $ 2.68\pm 0.31$ & $ 0.05\pm 0.05$ & $26.10\pm 0.03$ & $ 3.64$ & $ 3.93^{+ 0.40}_{- 0.33}$ & $ 6.6^{+ 0.8}_{- 0.7}$ &   ${\rm A_{14}}$, I, ${\rm A_{15}}$\\
1695-4527 & i2 & $ 3.570654 $ & $ -30.414659 $ & $ 1.33\pm 0.10$ & $ 0.11\pm 0.05$ & $26.22\pm 0.04$ & $ 1.47$ & $ 1.54^{+ 0.09}_{- 0.06}$ & $ 6.0^{+ 0.7}_{- 0.7}$ &   ${\rm A_{14}}$, I, ${\rm A_{15}}$\\
2549-3119 & i3 & $ 3.606222 $ & $ -30.386644 $ & $ 1.07\pm 0.08$ & $ 0.08\pm 0.05$ & $26.26\pm 0.04$ & $ 1.55$ & $ 1.65^{+ 0.08}_{- 0.06}$ & $ 5.8^{+ 0.7}_{- 0.6}$ &   ${\rm A_{14}}$, I, ${\rm A_{15}}$\\
1930-4181 & i4 & $ 3.580453 $ & $ -30.405043 $ & $>  2.36$ & $ 0.17\pm 0.09$ & $26.61\pm 0.06$ & $ 4.47$ & $ 4.90^{+ 0.43}_{- 0.36}$ & $ 6.8^{+ 0.8}_{- 0.7}$ &   ${\rm A_{14}}$, I, ${\rm A_{15}}$\\
2516-4570 & i5 & $ 3.604865 $ & $ -30.415839 $ & $ 0.92\pm 0.13$ & $ 0.06\pm 0.07$ & $26.67\pm 0.06$ & $ 2.84$ & $ 3.02^{+ 0.24}_{- 0.22}$ & $ 5.6^{+ 0.7}_{- 0.6}$ & \nodata \\
2348-3454 & i6 & $ 3.597834 $ & $ -30.395961 $ & $>  2.13$ & $ 0.27\pm 0.10$ & $26.79\pm 0.07$ & $ 3.31$ & $ 3.47^{+ 0.27}_{- 0.21}$ & $ 7.0^{+ 0.8}_{- 0.8}$ &   I, ${\rm A_{15}}$\\
2178-2458 & i7 & $ 3.590761 $ & $ -30.379409 $ & $ 0.96\pm 0.13$ & $-0.16\pm 0.09$ & $27.07\pm 0.08$ & $ 1.68$ & $ 1.78^{+ 0.12}_{- 0.09}$ & $ 5.9^{+ 0.6}_{- 0.7}$ &   I\\
2047-3526 & i8 & $ 3.585322 $ & $ -30.397958 $ & $>  1.83$ & $ 0.13\pm 0.14$ & $27.19\pm 0.11$ & $ 3.23$ & $ 3.76^{+ 0.36}_{- 0.34}$ & $ 6.8^{+ 0.7}_{- 0.8}$ &   ${\rm WZ_{14}}$, I, ${\rm A_{15}}$\\
\enddata
\tablenotetext{a}{For simplicity, the part that represents the cluster name is omitted. Examples of full IDs are HFF1C-2251-4556 and HFF1C-i1.}
\tablenotetext{b}{Short ID used only in this paper for simplification and clarity.}
\tablenotetext{c}{Total magnitude.}
\tablenotetext{d}{Median value and $1\sigma$ error of the magnification
factor from the MCMC posterior distribution.}
\tablenotetext{e}{${\rm A_{14}}$~=~\citet{atek14}, B~=~\citet{bradley14}, ${\rm WZ_{14}}$~=~\citet{zheng14}, I~=~\citet{ishigaki15}, ${\rm A_{15}}$~=~\citet{atek15}, C~=~\citet{coe15}.}
\end{deluxetable*}

\LongTables
\begin{deluxetable*}{llcclrcrrcl} 
\tabletypesize{\scriptsize}
\tablecaption{Dropout galaxy candidates at $z\sim 8$\label{tab:candidates8} 
in the four HFF cluster fields.}
\tablewidth{0pt}
\tablehead{
\colhead{ID\tablenotemark{a}} & \colhead{ID\tablenotemark{a,b}} & \colhead{R.A.} & \colhead{Decl.} & \colhead{$Y_{105}-J_{125}$} & \colhead{$J_{125}-J\!H_{140}$} & \colhead{$J\!H_{140}$\tablenotemark{c}} & \colhead{$\mu_{\rm best}$} & \colhead{$\mu$\tablenotemark{d}} & \colhead{$z_{\rm photo}$} & \colhead{Reference\tablenotemark{e}}
}
\startdata
\sidehead{HFF1C}
2508-2496 & Y1 & $ 3.604518 $ & $ -30.380467 $ & $ 1.17\pm 0.07$ & $ 0.04\pm 0.04$ & $25.92\pm 0.03$ & $ 1.31$ & $ 1.38^{+ 0.07}_{- 0.05}$ & $ 8.0^{+ 0.9}_{- 0.9}$ &   ${\rm L_{14}}$, ${\rm A_{14}}$, ${\rm WZ_{14}}$, I, ${\rm A_{15}}$, C\\
2481-2561 & Y2 & $ 3.603378 $ & $ -30.382255 $ & $ 1.27\pm 0.15$ & $ 0.14\pm 0.08$ & $26.64\pm 0.05$ & $ 1.40$ & $ 1.48^{+ 0.08}_{- 0.06}$ & $ 8.2^{+ 0.9}_{- 0.9}$ &   ${\rm WZ_{14}}$, I, ${\rm A_{15}}$, C\\
2306-3089 & Y3 & $ 3.596091 $ & $ -30.385833 $ & $ 1.30\pm 0.15$ & $ 0.06\pm 0.08$ & $26.68\pm 0.05$ & $ 2.04$ & $ 2.20^{+ 0.15}_{- 0.12}$ & $ 8.2^{+ 0.9}_{- 0.9}$ &   ${\rm WZ_{14}}$, I, ${\rm A_{15}}$, C\\
2555-2515 & Y4 & $ 3.606461 $ & $ -30.380996 $ & $ 1.12\pm 0.13$ & $-0.18\pm 0.09$ & $26.95\pm 0.06$ & $ 1.33$ & $ 1.40^{+ 0.07}_{- 0.05}$ & $ 8.0^{+ 0.8}_{- 0.9}$ &   ${\rm WZ_{14}}$, I, ${\rm A_{15}}$, C\\
2492-2561 & Y5 & $ 3.603859 $ & $ -30.382264 $ & $ 1.92\pm 0.32$ & $ 0.16\pm 0.10$ & $26.99\pm 0.07$ & $ 1.40$ & $ 1.48^{+ 0.08}_{- 0.06}$ & $ 8.4^{+ 0.9}_{- 0.9}$ &   ${\rm WZ_{14}}$, I, ${\rm A_{15}}$, C\\
2557-2513 & Y6 & $ 3.606576 $ & $ -30.380924 $ & $ 1.10\pm 0.22$ & $ 0.22\pm 0.13$ & $27.17\pm 0.08$ & $ 1.32$ & $ 1.39^{+ 0.07}_{- 0.05}$ & $ 7.9^{+ 0.9}_{- 6.1}$ &   ${\rm WZ_{14}}$, I, ${\rm A_{15}}$\\
2135-2432 & Y7 & $ 3.588980 $ & $ -30.378668 $ & $ 1.13\pm 0.18$ & $-0.19\pm 0.12$ & $27.30\pm 0.09$ & $ 1.69$ & $ 1.79^{+ 0.12}_{- 0.09}$ & $ 8.0^{+ 0.8}_{- 0.9}$ &   ${\rm WZ_{14}}$, I, ${\rm A_{15}}$, C\\
2495-2562 & Y8 & $ 3.603997 $ & $ -30.382304 $ & $ 1.05\pm 0.28$ & $ 0.13\pm 0.18$ & $27.61\pm 0.11$ & $ 1.40$ & $ 1.48^{+ 0.08}_{- 0.06}$ & $ 7.9^{+ 0.9}_{- 6.1}$ &   ${\rm WZ_{14}}$, I,  C\\
\enddata
\tablenotetext{a}{For simplicity, the part that represents the cluster name is omitted. Examples of full IDs are HFF1C-2508-2496 and HFF1C-Y1.}
\tablenotetext{b}{Short ID used only in this paper for simplification and clarity.}
\tablenotetext{c}{Total magnitude.}
\tablenotetext{d}{Median value and $1\sigma$ error of the magnification
factor from the MCMC posterior distribution.}
\tablenotetext{e}{${\rm L_{14}}$~=~\citet{laporte14}, ${\rm A_{14}}$~=~\citet{atek14}, ${\rm WZ_{14}}$~=~\citet{zheng14}, I~=~\citet{ishigaki15}, ${\rm L_{15}}$~=~\citet{laporte15}, ${\rm A_{15}}$~=~\citet{atek15}, C~=~\citet{coe15}, M~=~\citet{mcleod15}.}
\tablenotetext{f}{Selected by our two criteria for \idrop\ and \ydrop\ galaxies.}
\end{deluxetable*}

\LongTables
\begin{deluxetable*}{llcclrcrrcl} 
\tabletypesize{\scriptsize}
\tablecaption{Dropout galaxy candidates at $z\sim 9$\label{tab:candidates9} 
in the four HFF cluster fields.}
\tablewidth{0pt}
\tablehead{
\colhead{ID\tablenotemark{a}} & \colhead{ID\tablenotemark{a,b}} & \colhead{R.A.} & \colhead{Decl.} & \colhead{$Y\!J\tablenotemark{c} -J\!H_{140}$} & \colhead{$J\!H_{140}-H_{160}$} & \colhead{$H_{160}$\tablenotemark{d}} & \colhead{$\mu_{\rm best}$} & \colhead{$\mu$\tablenotemark{e}} & \colhead{$z_{\rm photo}$} & \colhead{Reference\tablenotemark{f}}
}
\startdata
\sidehead{HFF1C}
2481-2561\tablenotemark{g} & Y2\tablenotemark{g} & $ 3.603379 $ & $ -30.382255 $ & $ 0.78\pm 0.09$ & $-0.05\pm 0.07$ & $26.69\pm 0.05$ & $ 1.40$ & $ 1.49^{+ 0.08}_{- 0.06}$ & $ 8.2^{+ 0.9}_{- 0.9}$ &   ${\rm WZ_{14}}$, I, ${\rm A_{15}}$\\
2492-2561\tablenotemark{g} & Y5\tablenotemark{g} & $ 3.603859 $ & $ -30.382262 $ & $ 1.12\pm 0.18$ & $ 0.19\pm 0.09$ & $26.79\pm 0.06$ & $ 1.40$ & $ 1.48^{+ 0.08}_{- 0.06}$ & $ 8.4^{+ 0.9}_{- 0.9}$ &   ${\rm WZ_{14}}$, I, ${\rm A_{15}}$, M\\
2220-4053 & YJ3 & $ 3.592512 $ & $ -30.401486 $ & $>  1.21$ & $ 0.55\pm 0.27$ & $27.38\pm 0.15$ & $14.12$ & $14.37^{+ 1.22}_{- 0.90}$ & $ 9.6^{+ 1.0}_{- 7.1}$ &   AZ, I, O\\
\hline
\sidehead{HFF2C} 
1151-4540\tablenotemark{g} & Y1\tablenotemark{g} & $ 64.047984 $ & $ -24.081671 $ & $ 0.97\pm 0.10$ & $ 0.00\pm 0.06$ & $26.50\pm 0.04$ & $ 1.43$ & $ 1.44^{+ 0.04}_{- 0.04}$ & $ 8.4^{+ 0.9}_{- 0.9}$ &   ${\rm L_{15}}$, C, M\\
0939-5354\tablenotemark{g} & Y2\tablenotemark{g} & $ 64.039161 $ & $ -24.093184 $ & $ 1.27\pm 0.16$ & $ 0.06\pm 0.07$ & $26.55\pm 0.05$ & $ 1.39$ & $ 1.39^{+ 0.03}_{- 0.03}$ & $ 8.5^{+ 0.9}_{- 0.9}$ &   ${\rm L_{15}}$,  M\\
0901-5172\tablenotemark{g} & Y3\tablenotemark{g} & $ 64.037566 $ & $ -24.088113 $ & $ 1.06\pm 0.27$ & $-0.25\pm 0.19$ & $27.82\pm 0.14$ & $ 1.70$ & $ 1.71^{+ 0.06}_{- 0.06}$ & $ 8.3^{+ 1.0}_{- 0.9}$ &   ${\rm L_{15}}$, C, M\\
\hline
\sidehead{HFF4C} 
3358-4457 & YJ1 & $ 177.389950 $ & $ +22.412711 $ & $>  2.39$ & $ 0.19\pm 0.05$ & $25.80\pm 0.03$ & $17.51$ & $ 6.44^{+21.86}_{- 3.64}$ & $ 9.2^{+ 1.0}_{- 1.0}$ &   ${\rm WZ_{12}}$\\
4024-4492\tablenotemark{g} & Y2\tablenotemark{g} & $ 177.417700 $ & $ +22.413691 $ & $ 1.11\pm 0.23$ & $-0.15\pm 0.16$ & $27.44\pm 0.11$ & $ 1.63$ & $ 1.69^{+ 0.23}_{- 0.14}$ & $ 8.4^{+ 0.9}_{- 0.9}$ & \nodata \\
\enddata
\tablenotetext{a}{For simplicity, the part that represents the cluster name is omitted. Examples of full IDs are HFF1C-2481-2561 and HFF1C-Y2.}
\tablenotetext{b}{Short ID used only in this paper for simplification and clarity.}
\tablenotetext{c}{$Y\!J \equiv (Y_{105}+J_{125})/2$}
\tablenotetext{d}{Total magnitude.}
\tablenotetext{e}{Median value and $1\sigma$ error of the magnification
factor from the MCMC posterior distribution.}
\tablenotetext{f}{${\rm WZ_{12}}$~=~\citet{zheng12}, AZ~=~\citet{zitrin14}, ${\rm WZ_{14}}$~=~\citet{zheng14}, I~=~\citet{ishigaki15}, ${\rm L_{15}}$~=~\citet{laporte15}, ${\rm A_{15}}$~=~\citet{atek15}, C~=~\citet{coe15}, M~=~\citet{mcleod15}, O~=~\citet{oesch15}.}
\tablenotetext{g}{Selected by our two criteria for \ydrop\ and \yjdrop\ galaxies.}
\end{deluxetable*}

\end{appendix}

\end{document}